\documentclass[nofootinbib,a4paper,twocolumn,10pt]{revtex4}

\usepackage{amsmath}
\usepackage{amsfonts}
\usepackage{amssymb}
\usepackage{amsthm}
\usepackage{mathrsfs}
\usepackage{dsfont}
\usepackage{esint}
\usepackage{graphicx}
\usepackage{physics}

\DeclareMathAlphabet{\mathpzc}{OT1}{pzc}{m}{it}

	\newcommand{\PropTo}{\propto}
	\newcommand{\AsymEq}{\sim}
	\newcommand{\ApproxEq}{\approx}

	\newcommand{\V}[1]{\ensuremath{\boldsymbol{#1}}}			

	\newcommand{\mr}[1]{\mathrm{#1}}			

	\newcommand{\br}[1]{\left( #1 \right)}

	\newcommand{\of}[1]{\!\br{#1}}

	\newcommand{\sbr}[1]{( #1 )}
	\newcommand{\sbrr}[1]{[ #1 ]}
	
	\newcommand{\sof}[1]{\!\sbr{#1}}
	\newcommand{\soff}[1]{\!\sbrr{#1}}

	\newcommand{\Sum}[2]{\sum\limits_{#1}^{#2}}

	\newcommand{\Int}[3]{\int\limits_{#1}^{#2}\mr{d}#3\,}
	
	\newcommand{\sSum}[2]{\sum_{#1}^{#2}}

	\newcommand{\sInt}[3]{\int_{#1}^{#2}\mr{d}#3\,}

	\newcommand{\EA}[1]{\xpc{#1}}

	\newcommand{\xpc}[1]{\left\langle #1 \right\rangle}

	\newcommand{\sEA}[1]{\sxpc{#1}}
	\newcommand{\sVar}[1]{\mr{Var}\soff{#1}}
	
	\newcommand{\sxpc}[1]{\langle #1 \rangle}

	\newcommand{\Reals}{\ensuremath{\mathbb{R}} }

	\newcommand{\Complexs}{\ensuremath{\mathbb{C}} }

	\newcommand{\Nabla}{\V{\nabla}}

	\newcommand{\Landau}[1]{\mathpzc{O}\of{#1}}
	\newcommand{\sLandau}[1]{\mathpzc{O}\sof{#1}}

		\newcommand{\Max}[2]{\max\of{#1,#2}}

		\newcommand{\Sign}[1]{\mr{sgn}\of{#1}}
		\newcommand{\Id}{\mathds{1}}

		\newcommand{\BesselJ}[2]{J_{#1}\of{#2}}

\newcommand{\PsiDet}{\psi_\mathrm{d}}
\newcommand{\PsiIn}{\psi_\mathrm{in}}

\newcommand{\TEO}{\hat{U}}

\newcommand{\Ham}{\hat{H}}

\newcommand{\FDP}{F}

\newcommand{\FDA}{\varphi}

\newcommand{\TDP}{P_\text{det}}

\newcommand{\Detect}{\hat{D}}

\newcommand{\PSI}{\Psi}

\newcommand{\Cauchy}{\mathcal{C}}
\newcommand{\Bromwich}{\mathcal{B}}

\begin{document}
  \title{Non-Hermitian and Zeno limit of quantum systems under rapid measurements}
  \date{Manuscript of \today}
  \author{Felix Thiel}
  \affiliation{Department of Physics, Institute of Nanotechnology and Advanced Materials, Bar-Ilan University, Ramat-Gan 52900, Israel}
  \affiliation{Physikalisches Institut, Albert-Ludwigs-Universit\"at Freiburg, Hermann-Herder-Str.~3, 79104 Freiburg, Germany}
  \author{David A. Kessler}
  \affiliation{Department of Physics, Institute of Nanotechnology and Advanced Materials, Bar-Ilan University, Ramat-Gan 52900, Israel}
  \begin{abstract}
    We investigate in depth the relation between the first detection time of an isolated quantum system  that is repeatedly perturbed by strong local measurements with a large fixed frequency $1/\tau$, determining whether it is in some given state $\ket{\PsiDet}$, and the time of absorption to the same state of the same system with the added imaginary potential $2i\hbar\dyad{\PsiDet}/\tau$.
    As opposed to previous works, we compare directly the solutions of both problems in the small $\tau$, i.e., Zeno, limit.
    We  find a scaling collapse in
    $F(t)$ with respect to $\tau$ and compute the total detection probability as well as the moments of the first detection time probability density $F(t)$ in the Zeno limit.
     We show that both  solutions approach the same result in this small $\tau$ limit, as long as the initial state $\ket{\PsiIn}$ is not parallel to the detection state, i.e. as long as $\abs{\ip*{\PsiDet}{\PsiIn}} < 1$.
    However, when this condition is violated,  the small probability density to detect the state on time scales much larger than $\tau$ is precisely a factor of four different for all such times.
    We express the solution of the Zeno limit of both problems formally in terms of an electrostatic analogy. 
    Our results are corroborated with numerical simulations.
  \end{abstract}

  \maketitle

  \section{Introduction}
  \label{sec:Intro}
    Isolated quantum systems evolve unitarily according to the Schr\"odinger equation with a Hermitian 
    Hamiltonian until a measurement, obeying the collapse postulate, is performed \cite{Cohen-Tannoudji2009a, Braginsky1992a}.
    Recently, there has been increasing interest in repeatedly measured quantum systems \cite{Shikano2010a, Goenuelol2011a, Yi2011a, Gurvitz2017a, Mukherjee2018a, Rose2018a, Ashida2018a},
    where the unitary evolution is disrupted periodically.
    In the quantum first detection problem \cite{%
      Bach2004a, Krovi2006a, Krovi2006b, Krovi2007a, Varbanov2008a,
      Gruenbaum2013a, Bourgain2014a, Stefanak2008a, Dhar2015a, Dhar2015b, Lahiri2019a,%
      Sinkovicz2015a, Sinkovicz2016a, Friedman2017a, Friedman2017b, Thiel2018a, Thiel2018b,%
      Thiel2019a, Thiel2019b, Thiel2019c, Thiel2019d, Thiel2019f%
    }, a detector probes the system repeatedly as to whether it resides in a given target state $\ket{\PsiDet}$, or not.
    The quantity of interest is the (random) first detection time $T$, the time of the first successful detection attempt.
    This is a generalization of the time-of-arrival problem \cite{%
      Allcock1969a, Allcock1969b, Allcock1969c, Damborenea2002a, Galapon2005a, Echanobe2008a, Muga2008a, %
      Ruschhaupt2009a, Sombillo2014a, Sombillo2016a%
    } and the quantum analogue to the important first passage problem of statistical mechanics \cite{%
      Redner2007a, Benichou2011a, Benichou2015a, Hartich2018a, Hartich2019a%
    }.
    In the context of quantum information, it can also be seen as a protocol for quantum search \cite{%
      Grover1997a, Aaronson2003a, Bach2004a, Childs2004a, Magniez2011a, Chakraborty2016a, Li2017a%
    } or state transfer \cite{Kay2010a}.

    The detection protocol is the sequence of times $\{ t_1, t_2, t_3, \hdots \}$ at which the observer attempts to detect the system.
    In the stroboscopic detection protocol, the system is probed every $\tau$ time units, $t_n = n\tau$, 
    and $T$ can only assume an integer multiple of the detection period $\tau$.
    The detection protocol is a pragmatic way out of the problems with continuously observed quantum systems, i.e. the Zeno effect \cite{Misra1977a, Itano1990a, Elliott2016a, Schaefer2014a, Mueller2017a, Do2019a}.
    This latter describes the lock-down of quantum evolution that occurs when the system is rapidly measured.
    In the Zeno limit, when $\tau\to0$ and the detection frequency diverges, the system ``has no time'' to penetrate the detection space, before the measurement projects out that component from the evolving quantum state, making successful detection impossible.
    The first detection probability then vanishes, an effect that has been called the quantum Zeno paradox.
    
    The dynamics of this stroboscopic measurement protocol can be analyzed in terms of the non-Hermitian operator $(1- \dyad*{\PsiDet} ) e^{-i\Ham\tau/\hbar}$.
 The question then naturally arises how this stroboscopic dynamics is related to the non-Hermitian Hamiltonian, or optical potential, which has been introduced to model the loss of probability due to quantum transitions, or equivalently quantum traps~\cite{Pearlstein1972a,Krapivsky2014a}
    \begin{equation}
    \hat{H}_\textrm{NH} = \hat{H} - i\hbar \Gamma \dyad{\PsiDet}
    \label{eq:Schroedi}
    \end{equation}
    The connection between these two problems was already proposed by Allcock \cite{Allcock1969b}, who showed in the context of a measurement of whether the particle is in the region $x>0$  that $\Gamma=2/\tau$, and that reducing $\tau$ to increase temporal resolution leads to difficulties, later subsumed under the Zeno paradox rubric.
     This connection was
    later employed by Muga and co-workers who motivated the use of an optical potential to model stimulated photon emission from the system \cite{Damborenea2002a, Delgado2006a, Echanobe2008a, Muga2008a, Ruschhaupt2009a}.
     Schulman showed that $\Gamma = 2/\tau$ also in the case of a two-level system \cite{Schulman1998a}, and discussed the connection of the Zeno limit to continuous monitoring.
      Eq.~\eqref{eq:Schroedi}, which has been studied intensively in Ref.~\cite{Krapivsky2014a}, is important in its own right, because non-Hermitian terms appear when modeling the finite lifetime of certain energy states, when dealing with dissipative optical media, in quantum jump approaches, or in certain quantum transport models \cite{Moiseyev2011a,Ho1983a, Meystre1988a, Dalibard1992a, Gisin1992a, Buchleitner1994a, Plenio1998a, Brun2002a, Caruso2009a, Agliari2010a, Muelken2011a, Krapivsky2014a, Giusteri2015a, Novo2015a}.
     Non-Hermitian systems are also readily implemented experimentally \cite{Xu2017a, Rivet2018a, Xiao2019a, Lapp2019a, Li2019a}.
    It should be noted that the dissipation term in Eq.~\eqref{eq:Schroedi} can be derived from a system-bath coupling \cite{Redfield1957a}, but here our interest is in comparing and contrasting it to the stroboscopic measurement dynamics, particularly in the Zeno limit.
        
     Our primary goal in this work is to use the recently obtained renewal equation solution to the stroboscopic dynamics problem to examine the connection to the solution of the time-independent Schr\"odinger equation with the non-Hermitian Hamiltonian (NHH), Eq.~\eqref{eq:Schroedi}.
    One motivation for this is, given that the previous approaches use a perturbative approach to calculate the leading order non-Hermitian Hamiltonian of the stroboscopic approach, one might be concerned about the presence of nontrivial effects at long times.
    Another is just to see exactly how the two solutions are related.
    We shall show how to recover the identity of the two solutions in the Zeno limit, provided that the initial state is orthogonal to the detection state, in which case the dynamics is slow, with a time scale of order $1/\tau$.
    We then treat the case where the initial state is not orthogonal to the detection state.
    Then, in the Zeno limit, there is an initial fast transient, over a time scale of order $\tau$, in the NHH case.
    For the stroboscopic dynamics, the transient consists of the first measurement only, (which is also a time period of exactly $\tau$, after which the dynamics is slow.
     We then relate the post-transient dynamics of the two models, and see that when the initial state is {\em parallel} to the detection state, the slow dynamics differ by a factor of four in the decay statistics.
     In Section ~\ref{sec:Examples}, we exemplify these results through a number of specific models, namely the nearest-neighbor length-6 ring and the infinite line, as well as a random Hamiltonian.
    We then proceed to discuss this limiting Zeno solution in more detail, relating it to the solution of a particular electrostatic problem.
    This electrostatic problem defines all the quantities necessary for constructing the limiting Zeno solution.
    Sec.~\ref{sec:Return} shortly discusses the behavior close to the return problem, i.e. for initial states that are almost parallel to $\ket{\PsiDet}$.
    We close in Sec.~\ref{sec:SumDisc} with discussion and summary.
    Some additional details are given in the appendices.
    App.~\ref{app:Adiabatic} presents the adiabatic elimination of the fast mode in Eq.~\eqref{eq:Schroedi}, and App.~\ref{app:Lazy} discusses the limit $\tau\to\infty$ of very slow detectors for the non-Hermitian Schr\"odinger equation.
    App.~\ref{app:InfLine} presents some details of the infinite line calculations.
    Finally, App.~\ref{app:Numerics} explains how we obtained our numerical data.

  \section{Formal solutions to both problems}
  \label{sec:FormalSol}
    \subsection{The non-Hermitian Schr\"odinger equation}
      We first review the solution of the continuous-time problem starting with the non-Hermitian Schr\"odinger equation \eqref{eq:Schroedi}, following closely the derivation in Ref.~\cite{Krapivsky2014a}.
      Denote the solution to Eq.~\eqref{eq:Schroedi} with initial condition $\ket{\psi(t=0)} = \ket{\PsiIn}$ by $\ket{\psi(t)}$.
      The squared norm of this state is the remaining probability in the system, the survival probability.
      Its negative derivative is the probability density function (pdf) of detection/dissipation times that equals
      \begin{align}
        \FDP^\PSI(t) 
        = & \nonumber
        - \dv{t} \ip{\psi(t)}{\psi(t)}
        \\ = & \nonumber
        - \qty(\dv{t} \bra{\psi(t)}) \ket{\psi(t)}
        - \bra{\psi(t)} \dv{t} \ket{\psi(t)}
        \\ = & \nonumber
        - \ev*{\qty[
          \frac{i}{\hbar}\Ham 
          - \frac{i}{\hbar}\Ham 
        - 2\Gamma\dyad{\PsiDet}
        ]}{\psi(t)}
        \\ = &
        2\Gamma\abs*{\ip*{\PsiDet}{\psi(t)}}^2      
        =:
        2\Gamma\abs*{\PSI(t)}^2       
        ,
      \label{eq:DefFDPPsi}
      \end{align}
      where we introduced the overlap of the solution with the detection state $\PSI(t) := \ip*{\PsiDet}{\psi(t)}$,
      which is the only piece of $\ket{\psi(t)}$ that we actually need.
      Hereafter $\PSI(t)$ is called ``the wave function''.
      We place a sub- or superscript `$\PSI$' to quantities derived from the NHH framework.
      We apply a Laplace transform to Eq.~\eqref{eq:DefFDPPsi},
      for which we use the Laplace partners $\PSI(s) := \sInt{0}{\infty}{t} e^{-st} \PSI(t)$ and 
      $\PSI(t) = \sInt{\Bromwich}{}{s} e^{st} \PSI(s) / (2\pi i)$:\footnote{
        We use the same symbol for functions in original and image domain.
        In our convention, the functions are identified by their arguments.
      }
      \begin{align}
        \FDP^\PSI(s)
        := &
        \Int{0}{\infty}{t} e^{-st} \FDP^\PSI(t)
        =
        2\Gamma 
        \int\limits_\Bromwich\frac{\dd \sigma}{2\pi i} 
        \PSI^*(s-\sigma)
        \PSI(\sigma)
        .
      \label{eq:FDPPsiLaplace}
      \end{align}
      Here, $f^*(z) := [f(z^*)]^*$ and $z^*$ is the complex conjugate.
      The integration contour is the Bromwich path $\Bromwich := \{ 0^+ + i \omega | \omega \in \Reals \}$ 
      that lies directly to the right of the imaginary axis.
      To obtain Eq.~\eqref{eq:FDPPsiLaplace}, we used that $[\PSI(t)]^*$ transforms to $\PSI^*(s)$ and that products in the time domain become convolutions in the Laplace domain.
      Furthermore, we have $\Re{s} > \Re{\sigma} \ge 0$.
      It is important to note that the complex contour integral in Eq.~\eqref{eq:FDPPsiLaplace} only picks up the residues of $\PSI(\sigma)$,
      but not those of $\PSI^*(s-\sigma)$.

      In addition to the pdf of $T$, we are also interested in the total detection probability $\TDP$
      and in $T$'s (conditional) moments:
      \begin{equation}
        \TDP
        :=
        \Int{0}{\infty}{t} \FDP(t)
        \qc
        \EA{T^m}
        :=
        \frac{1}{\TDP} \Int{0}{\infty}{t} t^m \FDP(t)
        .
      \label{eq:DefTDPMoments}
      \end{equation}
      $\TDP$ is the fraction of experimental runs in which the detector finds something at all.
      It is also the normalization of the first detection time pdf.
      $\EA{T^m}$ is the $m$-th moment of $T$ provided that the system was detected at all.
      Obviously, Eq.~\eqref{eq:DefTDPMoments} holds for the stroboscopic framework as well, whence the lack of superscripts.

      Instead of computing these quantities in the time domain, we can also obtain them from the Laplace quantity $\PSI(s)$,
      by using a version of Parseval's theorem:
      \begin{align}
      \label{eq:DefTDPSchroedi}
        \TDP^\PSI
        = &
        2\Gamma   
        \int\limits_\Bromwich\frac{\dd s}{2\pi i}
        \PSI^*(-s) \PSI(s)
        \\
        \EA{T^m}^\PSI
        = &
        \frac{2 \Gamma}{\TDP^\PSI}
        \int\limits_\Bromwich\frac{\dd s}{2\pi i}
        \PSI^*(-s) 
        \qty( - \tfrac{\dd}{\dd s})^m
        \PSI(s)
        .
      \label{eq:DefMomentSchroedi}
      \end{align}
      Here we have expressed the integral as a Laplace transform at $s=0^+$ and proceeded by using $t f(t) \mapsto - \tfrac{\dd}{\dd s}f(s)$.
      Let us now determine $\PSI(s)$.

      To do so, we apply a Laplace transform to Eq.~\eqref{eq:Schroedi}.
      Here we explicitly use the initial condition $\ket{\PsiIn}$, because $(\dd / \dd t) f(t) \mapsto sf(s) - f(t=0)$.
      We write $\ket{\psi(s)} := \sInt{0}{\infty}{t} e^{-st} \ket{\psi(t)}$.
      \begin{equation}
        i \hbar [ s \ket{\psi(s)} - \ket{\PsiIn} ]
        =
        \Ham \ket{\psi(s)}
        -
    i\Gamma\hbar  
         \ket{\PsiDet}
        \ip{\PsiDet}{\psi(s)}
        .
      \label{eq:}
      \end{equation}
      The equation is rearranged
      \begin{equation}
        \ket{\psi(s)}
        =
        \qty[ s + \frac{i}{\hbar} \Ham ]^{-1}
        \qty[
          \ket{\PsiIn}
          -
         \Gamma 
           \ket{\PsiDet}
          \ip{\PsiDet}{\psi(s)}
        ]
        ,
      \label{eq:}
      \end{equation}
      multiplied with $\bra{\PsiDet}$ from the left and solved
      for $\ip{\PsiDet}{\psi(s)} = \PSI(s)$:
      \begin{equation}
        \PSI(s)
        =
        \frac{
          \mel{\PsiDet}{\frac{1}{s + \frac{i}{\hbar}\Ham}}{\PsiIn}
        }{
          1
          + \Gamma 
          \mel{\PsiDet}{\frac{1}{s + \frac{i}{\hbar}\Ham}}{\PsiDet}
        }
        =:
        \frac{ v_\PSI(s)}{1 + \Gamma
        u_\PSI(s)}
        ,
      \label{eq:DefPsi}
      \end{equation}
      where we have abbreviated:
      \begin{equation}
        u_\PSI(s)
        := 
        \mel{\PsiDet}{\frac{1}{s + \frac{i}{\hbar}\Ham}}{\PsiDet}
        \qc 
        v_\PSI(s)
        := 
          \mel{\PsiDet}{\frac{1}{s + \frac{i}{\hbar}\Ham}}{\PsiIn}
      \label{eq:DefResolventSchroedi}
      \end{equation}
      $u_\PSI(s)$ and $v_\PSI(s)$ are thus the diagonal and off-diagonal, respectively, matrix elements of the Hamiltonian's resolvent.
      In the return problem, when the initial and detection states are identical, $v_\PSI(s) = u_\PSI(s)$, but for the transition problem, where these two states are different, the two quantities are different.
      Thus, specification of the Hamiltonian, together with the initial and detection states, yields the two functions $u_\PSI(s)$ and $v_\PSI(s)$.
      From these one finds, via Eq.~\eqref{eq:DefPsi}, $\PSI(s)$, which in turn gives the pdf $F^\PSI(s)$ and all moments via integration.

    \subsection{The stroboscopic detection protocol}
      We now review the solution of the stroboscopic detection protocol.
     Here, the detection time can only assume integer multiples of $\tau$.
      Consequently the (quasi-continuous-time) pdf of $T$ must be a comb of delta functions, that is $\FDP^\FDA(t) = \sSum{n=1}{\infty} \abs{\FDA_n}^2 \delta(t - n\tau)$.
      The first detection amplitude's squared modulus $\abs*{\FDA_n}^2$ gives the probability that the $n$-th 
      detection attempt is the first successful one.
      $\FDA_n$ is given by \cite{Dhar2015a, Friedman2017b}:
      \begin{equation}
        \FDA_n
        =
        \mel*{\PsiDet}{\TEO(\tau) [ (\Id - \Detect)\TEO(\tau) ]^{n-1}}{\PsiIn}
        ,
      \label{eq:DefFDA}
      \end{equation}
      where $\Detect = \dyad{\PsiDet}$.
      According to \cite{Friedman2017b}, it can alternatively be obtained from a renewal equation via:
      \begin{equation}
        \FDA_n
        =
        \mel*{\PsiDet}{[\TEO(\tau)]^n}{\PsiIn}
        -
        \Sum{m=1}{n-1} 
        \mel*{\PsiDet}{[\TEO(\tau)]^m}{\PsiDet}
        \FDA_{n-m}
        .
      \label{eq:QuantumRenewal}
      \end{equation}
      The generating function $\FDA(z) := \sSum{n=1}{\infty} \FDA_n z^n$ can be obtained from this equation
      by multiplying it with $z^n$ and then summing over all $n$ from one to infinity.
      This recovers the definition of $\FDA(z)$ on the left-hand side.
      The convolution on the right hand side becomes a product in the $z$-domain,
      the terms $[\TEO(\tau)]^n$ are gathered in a geometric series and the equation is solved for $\FDA(z)$~\cite{Friedman2017b}:
      \begin{align}
        \FDA(z)
        =
        \frac{
          \mel*{\PsiDet}{\frac{z \TEO(\tau)}{1 - z\TEO(\tau)}}{\PsiIn}
        }{
          \mel*{\PsiDet}{\frac{1}{1 - z\TEO(\tau)}}{\PsiDet} 
        }
        =:
        \frac{v_\FDA(z) - \ip{\PsiDet}{\PsiIn}}{u_\FDA(z)}
        .
      \label{eq:DefGenFunc}
      \end{align}
      Analogously to Eq.~\eqref{eq:DefPsi}, we have defined:
      \begin{equation}
        u_\FDA(z)
        :=
        \mel*{\PsiDet}{\frac{1}{1 - z\TEO(\tau)}}{\PsiDet} 
        , \;
        v_\FDA(z) 
        := 
          \mel*{\PsiDet}{\frac{1}{1 - z\TEO(\tau)}}{\PsiIn}
        .
      \label{eq:DefResolventStrobo}
      \end{equation}
      $u_\FDA(z)$ is the (slightly differently defined) resolvent of the evolution operator, 
      and $v_\FDA(z)$ is equal to $u_\FDA(z)$  in the return problem.
      The sub- or superscript `$\FDA$' denotes quantities derived from the stroboscopic detection protocol.

      This generating function reappears in the Laplace transform $\FDP^\FDA(s)$, which is the generating function of the product
      $\FDA_n^* \FDA_n$ evaluated at $z=e^{-s\tau}$. 
      Again using that products in the time domain become convolutions in the $z$-domain, we find:
      \begin{equation}
        F^\FDA(s)
        =
        \Sum{n=1}{\infty} \FDA_n^* \FDA_n e^{-n s\tau}
        =
        \ointctrclockwise\limits_\Cauchy
        \frac{\dd z}{2 \pi i z}
        \FDA^*\qty(\tfrac{e^{-s\tau}}{z}) \FDA(z)
        ,
      \label{eq:FDPPhiLaplace}
      \end{equation}
      a result analogous to Eq.~\eqref{eq:FDPPsiLaplace}.
      Here the integration follows the Cauchy contour $\Cauchy = \{ e^{i\omega+0^-} | \omega \in [-\pi, \pi] \}$ just inside the unit circle.
      To compute the total detection probability and the moments, we use $n f_n \mapsto z \tfrac{\dd}{\dd z} f(z)$ and find:
      \begin{align}
        \label{eq:TDPStrobo}
        \TDP^\FDA
        = &
        \ointctrclockwise\limits_\Cauchy
        \frac{\dd z}{2 \pi i z}
        \FDA^*\qty(\tfrac{1}{z}) \FDA(z)
        \\
        \EA{T^m}^\FDA
        = &
        \frac{1}{\TDP^\FDA}
        \ointctrclockwise\limits_\Cauchy
        \frac{\dd z}{2 \pi i z}
        \FDA^*\qty(\tfrac{1}{z}) 
        (z \tfrac{\dd }{\dd z} )^m
        \FDA(z)
        .
        \label{eq:MomentStrobo}
      \end{align}
      Knowledge of the generating function $\FDA(z)$ is sufficient to obtain all quantities pertaining to the stroboscopic detection protocol.

  \section{Small $\tau$ limit of the stroboscopic detection protocol}
  \label{sec:Equivalence}
    Starting from Eq.~\eqref{eq:FDPPhiLaplace}, we now demonstrate how (and under what conditions) the solution of the non-Hermitian 
    Schr\"odinger equation $\FDP^\PSI(t)$ emerges from $\FDA_n$ in the limit of small $\tau$ and large $n$, such that $t=n\tau$ remains constant.
    The limit $n\to\infty$ is most conveniently taken in the $z$-domain, where it corresponds to $\abs{z}\to1^-$, i.e. approaching closer and closer to the unit circle.

    The two key steps are the variable change $z=e^{-s\tau}$ and the asymptotic equality
    \begin{equation}
      \frac{1}{1 - e^{-x\tau}}
      =
      \frac{1}{x\tau} 
      + \frac{1}{2} + \Landau{\tau}
      ,
    \label{eq:}
    \end{equation}
    which is used to replace:
    \begin{equation}
      \mel*{\psi}{\frac{1}{1 - e^{-\tau(s + i \Ham/\hbar)}}}{\psi'}
      =
      \frac{1}{\tau} \mel*{\psi}{\frac{1}{s + \frac{i}{\hbar} \Ham}}{\psi'}
      + \frac{\ip{\psi}{\psi'}}{2} + \Landau{\tau}
      ,
    \label{eq:}
    \end{equation}
    for two arbitrary states $\ket{\psi}$ and $\ket{\psi'}$.
    In terms of the previously defined functions, this means:
    \begin{align}
      \label{eq:ResCorrespondence}
      u_\FDA(e^{-s\tau})
      \AsymEq &
      \frac{1}{\tau} u_\PSI(s) + \frac{1}{2}
      \\
      v_\FDA(e^{-s\tau})
      \AsymEq &
        \frac{1}{\tau} v_\PSI(s) + \frac{\ip{\PsiDet}{\PsiIn}}{2}
      \label{eq:NCorrespondence}
      .
    \end{align}
    
    At this point, it is convenient to assume that the initial state has no overlap with the detection state, $\ip{\PsiDet}{\PsiIn}=0$.
 We will in the next section deal with the more general problem.
    Plugging these results into the generating function $\FDA(z=e^{-s\tau})$ gives us the relation
    \begin{equation}
      \varphi(e^{-s\tau})
      \AsymEq
      \frac{
        \frac{2}{\tau}
        \mel{\PsiDet}{\frac{1}{s + \frac{i}{\hbar}\Ham}}{\PsiIn} }{
        1 + \frac{2}{\tau}\mel{\PsiDet}{\frac{1}{s + \frac{i}{\hbar}\Ham}}{\PsiDet}
      }
      = 
      \frac{2}{\tau} \PSI(s)
      .
    \label{eq:FDAApprox}
    \end{equation}
    The result is, up to the factor $2/\tau$, just the wave function $\PSI(s)$ of the NHH problem, with the identification $\Gamma = 2/\tau$.
    $\PSI(s)$ appears in Eq.~\eqref{eq:FDPPhiLaplace} after the change of variables $z = e^{-\sigma\tau}$
    with $\dd z = - \tau z \dd \sigma$.
    This changes the Cauchy contour to the ``proto-Bromwich path'' $\Bromwich_\tau := \{ 0^+ + i \omega | \omega \in [-\pi/\tau, \pi/\tau]\}$,
    which converges to the inverse Laplace transform's Bromwich path as $\tau\to0$.
    One then finds:
    \begin{equation}
      \FDP^\FDA(s)
      \AsymEq
      \frac{4}{\tau}
      \int\limits_{\Bromwich_\tau}
      \frac{\dd \sigma}{2 \pi i}
      \PSI^*(s-\sigma) \PSI(\sigma)
      ,
    \label{eq:FDPPhiApprox}
    \end{equation}
    recovering Eq.~\eqref{eq:FDPPsiLaplace} after the replacement \eqref{eq:FDAApprox}.
    We have thus shown that $\FDP^\FDA(t) \AsymEq \FDP^\PSI(t)$.
    Integration of this relation immediately yields $\EA{T^m}^\FDA \AsymEq \EA{T^m}^\PSI$ as well, so that as expected the stroboscopic protocol does reduce to the NHH formalism in the Zeno limit $\tau \ll 1$, conditioned however on the orthogonality of the detection state to the initial state.

  \section{Small $\tau$ limit for overlapping initial and detection states}
  \label{sec:MagicFactor}
    Before we turn to analyze the more general situation $\ip*{\PsiIn}{\PsiDet} \ne 0$, let us note that when initial and detection state are orthogonal, the typical time scale on which $F(t)$ decays is slow, that is of order $\sLandau{\Gamma} = \sLandau{1/\tau}$.
    We shall show this explicitly in Sec.~\ref{sec:Ultimate}.
    This somewhat non-intuitive result is a consequence of the quantum Zeno effect.
    The large magnitude $\Gamma = 2/\tau$ of the optical potential on $\ket{\PsiDet}$ results in effective reflection of the wave function off the detection state.
    Thus the overlap of the wave function on the detection state $\PSI(t) = \ip*{\PsiDet}{\psi(t)}$ is always small of order $\sLandau{1/\Gamma^2} = \sLandau{\tau^2}$, resulting in a slow $\sLandau{1/\Gamma} = \sLandau{\tau}$ decay of probability.

    This situation obviously cannot hold in the case when  $\ip{\PsiDet}{\PsiIn}\ne 0$, since the overlap initially is of order unity.
    What happens in this case is that the overlap rapidly decays to its typically small value over the short time scale $\sLandau{\tau} = \sLandau{1/\Gamma}$.
    Thus, in the small $\tau$ limit, we can map the  $\ip{\PsiDet}{\PsiIn}\ne 0$ case to an equivalent  $\ip{\PsiDet}{\PsiIn}= 0$ case after this short transient.  
  
    Similarly, in the stroboscopic detection protocol, when  $\tau \ll 1$ and $\ip{\PsiDet}{\PsiIn}\ne 0$ there is an initial transient after which the problem reduces to that of an equivalent
   $\ip{\PsiDet}{\PsiIn}= 0$ case.
    The only difference with the NHH case is that the transient only lasts until the first detection attempt, as opposed to decaying exponentially.
   
    So, to proceed, we first analyze the NHH case by considering the survival probability after a very short time.
    Clearly, when $\Gamma$ is very large, and the initial state is equal to the detection state, most of the amplitude will be absorbed shortly after preparation.
    As we will demonstrate in detail below, the non-Hermitian Hamiltonian of Eq.~\eqref{eq:Schroedi} has exactly one fast mode $\ket{\psi_f}$ with eigenvalue $-i\hbar\Gamma + \hbar \omega_0 + \sLandau{1/\Gamma}$, where $\omega_0 = \ev*{\Ham}{\PsiDet}/\hbar$.
    The fast mode is given to $\Landau{1/\Gamma}$ by 
    \begin{equation}
      \ket{\psi_f} 
      = 
    \left[ \Id + \frac{i}{\hbar\Gamma} (\Id - \Detect) \Ham\right]
      \ket{\PsiDet}
      .
    \label{eq:FastMode}
    \end{equation}
    Over small times $1/\Gamma \ll t \ll1$, all of the wave function's overlap with the fast mode will be lost.
    In the case of partial but not complete overlap between initial and detection states,  $0 < \abs*{\ip*{\PsiDet}{\PsiIn}}^2 < 1$, the solution $\ket{\psi(t)}$ of Eq.~\eqref{eq:Schroedi} at small times is given to leading order by
    \begin{align}
      \ket{\psi(t)}
      \AsymEq & \nonumber
      e^{- t(\Gamma + i \omega_0)}
      \ket{\psi_f} \ip*{\psi_f}{\PsiIn}
      +
      (\Id - \dyad{\psi_f})\ket{\PsiIn}
      \\ \AsymEq &
      (\Id - \Detect)\ket{\PsiIn}
    \end{align}
    Thus, after the fast transient dies away, $\ket{\PsiIn^\text{eff}}_\PSI = (\Id - \Detect )\ket{\PsiIn}$, and the survival probability is $S_\PSI = 1 - |\ip{\PsiDet}{\PsiIn}|^2$.
 
    The situation is different for full overlap $\abs{ \ip*{\PsiDet}{\PsiIn} }^2 = 1$.
    Here, almost all of the probability is depleted during the transient, and the post-transient ``initial'' wave function is 
    \begin{equation}
      \ket{\PsiIn^\text{eff}}_\PSI
      = 
      \frac{1}{i\hbar\Gamma} 
      (\Id - \Detect) 
      \Ham\ket{\PsiDet}
    \label{eq:PsiPost}
    \end{equation}
    whose magnitude gives the tiny 
    survival probability $S_\PSI$:
    \begin{equation}
      S_\PSI
      =
      \frac{1}{\hbar^2 \Gamma^2} \bra{\PsiDet}\Ham(\Id - \Detect)\Ham\ket{\PsiDet}
      \AsymEq
      \frac{1}{4}
      \frac{\tau^2}{\tau_\text{Z}^2} 
      ,
    \label{eq:FastRemoved1}
    \end{equation}
    where we replaced $\Gamma = 2/\tau$ and $\tau_\text{Z}^2 = \hbar^2 / \ev*{\Ham(\Id-\Detect)\Ham }{\PsiDet}$ is the Zeno time \cite{Facchi2008a}.    
    
    We now turn to the stroboscopic detection protocol.
    For partial overlap, to leading order, the detection probability in the first detection attempt is 
    $\abs{\FDA_1}^2 = \abs{\ip{\PsiDet}{\PsiIn}}^2$ and the survival probability is $S_\FDA = 1 - \abs{\FDA_1}^2 =  1 - |\ip{\PsiDet}{\PsiIn}|^2$.
    This is just the post-transient survival probability of the NHH problem, and so both problems map on to calculating the subsequent survival of the post-transient state $(\Id - \Detect)\ket{\PsiIn}$.
    As we have seen above, this survival probability is identical to leading order in the two problems, the two problems are thus seen to yield identical results also for partial overlap.
    
    For complete overlap, however, things are very different.
    For the stroboscopic protocol, we have that the first detection attempt is almost surely successful, and the probability of survival is due to the small probability that transferred to other states in the small time $\tau$.
    The wave function immediately after  the first measurement is 
    \begin{equation}
      \ket{\PsiIn^\text{eff}}_\FDA
      = 
      (\Id - \Detect) e^{-i\frac{\tau \Ham}{\hbar}} \ket{\PsiDet}
      \AsymEq
      -i\frac{\tau}{\hbar} (\Id-\Detect)\Ham \ket{\PsiDet}
    \label{eq:PhiPost}
    \end{equation}
    and the survival probability is then
    \begin{equation}
      S_\FDA
      =
      \frac{\tau^2}{\hbar^2} \norm*{
        (\Id - \Detect) \Ham
        \ket{\PsiDet}
      }^2
      \AsymEq
      \frac{\tau^2}{\tau_\text{Z}^2} 
      .
    \label{eq:FastRemoved2}
    \end{equation}
    Thus, while the form of the post-transient wave function Eq.~\eqref{eq:PhiPost} is the same as that for the NHH, Eq.~\eqref{eq:PsiPost}, there is a factor of two difference, under the identification $\Gamma=2/\tau$, along with a factor four discrepancy between Eqs.~\eqref{eq:FastRemoved1} and \eqref{eq:FastRemoved2}.
    This difference stems from the quantitative difference in the transient dynamics of the return problem in the two models.
    It arises from the fact that the evolution under stroboscopic detection is free until time $\tau$, when the first amplitude is removed, while the evolution under Eq.~\eqref{eq:Schroedi} is dissipative from the very beginning, such that less amplitude survives.

    To relate the total detection probability of both formalisms for complete overlap, we write $\TDP^\FDA = \sSum{n=2}{\infty} \abs*{\FDA_n}^2 + 1 - S_\FDA \AsymEq 4 [ \TDP^\PSI -(1 - S_\PSI) ] + 1 - S_\FDA$.
    Since the survival probabilities are $\sLandau{\tau^2}$, we can neglect them for $\TDP$.
    For the moments, these initial discrepancies are not relevant.
    We find:
    \begin{align}
      \TDP^\FDA 
      \AsymEq 
      4 \TDP^\PSI - 3 
      \qc
      \EA{T^m}^\FDA 
      \AsymEq 
      \frac{4 \TDP^\PSI}{4 \TDP^\PSI - 3} 
      \EA{T^m}^\PSI
      ,
    \label{eq:TDPMomentEquiv}
    \end{align}
    for $\abs{\ip*{\PsiDet}{\PsiIn}}=1$.
    Any other initial condition gives $\TDP^\FDA \AsymEq \TDP^\PSI$ and $\sEA{T^m}^\FDA \AsymEq \sEA{T^m}^\PSI$.
    We find from Eq.~\eqref{eq:TDPMomentEquiv} that $\TDP^\FDA =1 \Leftrightarrow \TDP^\PSI =1$.
    Furthermore, the equivalence of the total detection probability must obviously break down when $\TDP^\PSI < 3/4$, which defines a critical upper limit for $\tau$.

    Note that the above expressions for the survival probabilities could be made coincident, if $\hbar/\tau$ is used as the optical potential strength in Eq.~\eqref{eq:Schroedi}.
    Then, however, all remaining aspects of $\FDP^\PSI(t)$ are not simply related to $\FDP^\FDA(t)$ as the following numerical investigations will show.

  \section{Examples}
  \label{sec:Examples}
    We demonstrate our results in three models.
    The first is the tight-binding model on the benzene ring, i.e. a ring with six sites:
    \begin{equation}
      \Ham_B
      := 
      -\gamma \Sum{x=0}{5} 
      \big[
        \dyad{x}{x-1}
        +\dyad{x}{x+1}
      \big]
      ,
    \label{eq:BenzHam}
    \end{equation}
    with periodic boundary conditions, such that $\ket{x+6} = \ket{x}$.
    $\gamma$ is the hopping energy that determines the width of the spectrum. 
    $\Ham_B$ has four energy levels $-2\gamma$, $-\gamma$, $\gamma$, $2\gamma$.
    The second system is a random $32\times32$-dimensional Hamiltonian $\Ham_R$ taken from the Gaussian unitary ensemble.
    This Hamiltonian's lack of symmetries, its random energy levels and eigenstates demonstrate that our results are not specific to any particular model. 
    Nevertheless, we rescaled the spectrum of $\Ham_R$ such that it lies between $-2\gamma$ and $2\gamma$.
    This way, the time scales of both systems are comparable.
    For each of the Figures \ref{fig:FDP} and \ref{fig:RandStat} one single sample matrix was used.\footnote{For Fig.~\ref{fig:RandStat} convergence of the stroboscopic data can be problematic. Therefore, we chose a GUE-matrix for which this issue is not so severe, see App.~\ref{app:Numerics}. }
    Finally, we also present a system with a continuous spectrum, the tight-binding model on the infinite line:
    \begin{equation}
      \Ham_L
      :=
      -\gamma
      \Sum{x=-\infty}{\infty} \big[
        \dyad{x}{x-1} 
        + \dyad{x}{x+1}
      \big]
      ,
    \label{eq:InfLineHam}
    \end{equation}
    whose spectrum again lies between $-2\gamma$ and $2\gamma$, but is not discrete as before.
    In our figures, we measure $\tau$ in units of $\hbar/\gamma$.
    The shortest time scale for all three Hamiltonians is given by the width of the energy spectrum: $\hbar / (4\gamma)$.
    All $\tau$ values have to be compared to this shortest system time scale $0.25 \hbar/\gamma$.

    For the benzene ring and the infinite line, the detection state was chosen to be a position eigenstate.
    For the random Hamiltonian, we did the same in the sampled basis of the matrix, such that:
    \begin{equation}
      \ket*{\PsiDet^{B,R,L}} = \ket{0}
      .
    \label{eq:DefPsiDetEx}
    \end{equation}
    We investigated a total of four initial states for the benzene ring:
    \begin{align}
      \ket*{\PsiIn^{B,1}} = \ket{1}
      \qc  & \nonumber
      \ket*{\PsiIn^{B,2}} = \ket{3},
      \\
      \ket*{\PsiIn^{B,3}} = \Sum{x=0}{5} \frac{\ket{x}}{\sqrt{6}}
      \qc &
      \ket*{\PsiIn^{B,4}} = \ket{0}
      .
    \label{eq:BenzPsiIn}
    \end{align}
    The last one is the detection state; the first one does not yield a unit total detection probability.
    The third is an eigenstate of the Hamiltonian.
    For the random Hamiltonian, we explored two initial states:
    \begin{equation}
      \ket*{\PsiIn^{R,1}} = \ket{1}
      \qc
      \ket*{\PsiIn^{R,2}} = \ket{0} = \ket{\PsiDet}
      ,
    \label{eq:RandPsiIn}
    \end{equation}
    so that the second initial state yields the return problem.
    Different position eigenstates have been chosen as initial states for the infinite line.

    App.~\ref{app:Numerics} explains in detail how the numerical data was obtained.
    Throughout this article's figures, we stick to the following color code:
    Data for stroboscopic detection is given by gray squares, data from the non-Hermitian Schr\"odinger equation 
    is depicted as blue crosses ($+$), corrected non-Hermitian data is given by 
    orange triangles ($\triangle$).
    Finally, data from the Zeno approximation (see below) is depicted by green circles ($\circ$).

    \subsection{Probability density functions}
      \begin{figure}
        \includegraphics[width=0.99\columnwidth]{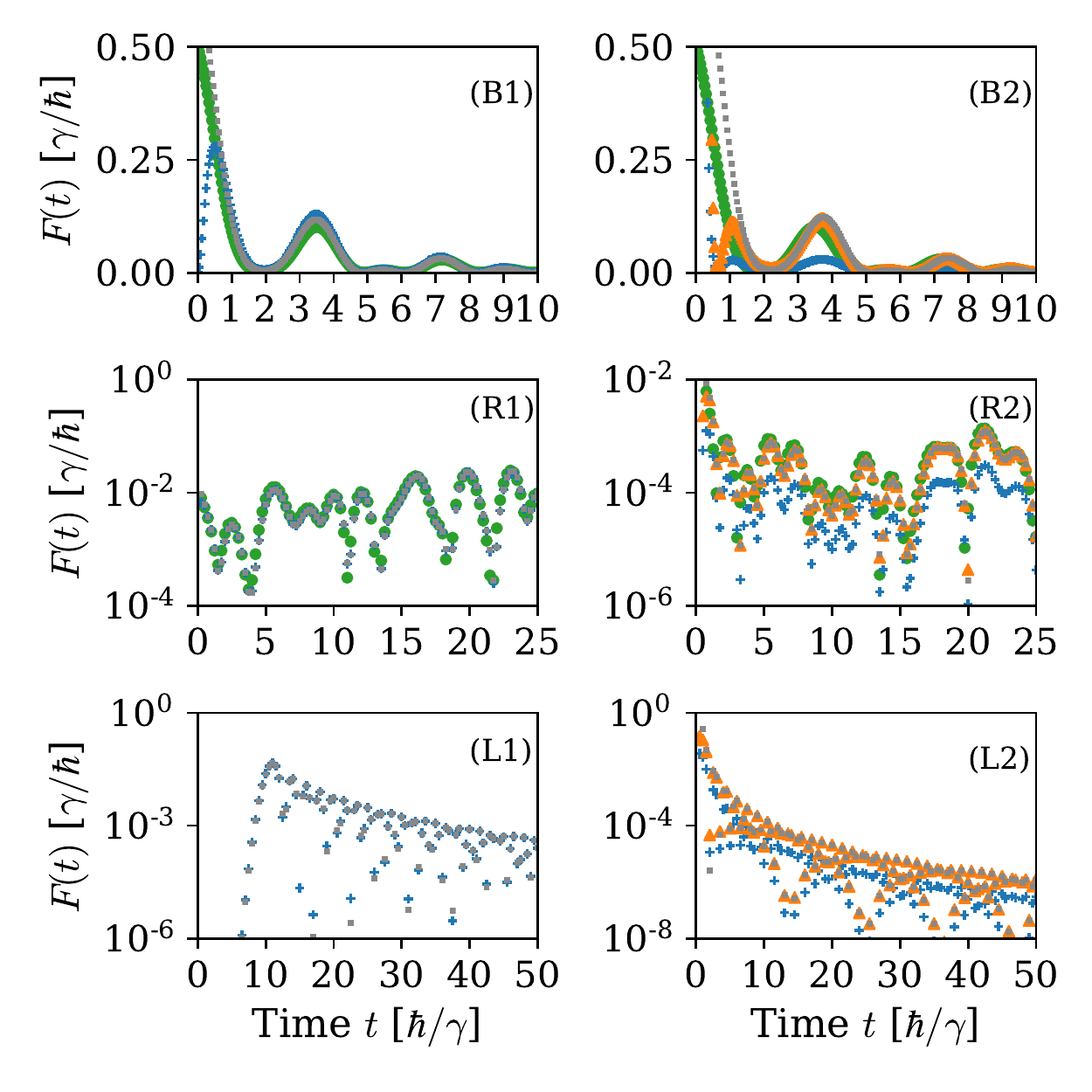}
        \caption{
          Comparison of $\FDP^\FDA(t)$ (gray squares), $\FDP^\PSI(t)$ (blue $+$), the Zeno approximation [green $\circ$, Eq.~\eqref{eq:FDPZenoStrobo}, not for (L)], and the corrected non-Hermitian approach with an additional factor four (orange $\triangle$, only right column).
          For the benzene ring (B), the random Hamiltonian (R) and on the infinite line (L).
          Detection is performed at the origin $\ket{\PsiDet} = \ket{0}$.
          Detection period is given by $\tau = 0.5 \hbar/\gamma$ (B,L) or $\tau = 0.25 \hbar/\gamma$ (R).
          Initial states are $\ket*{\psi_{\text{in},1}^{B,R}}$ and $\ket*{\PsiIn^L} = \ket{20}$ for the left column (1) and $\ket{\PsiDet}$ for the right column (2).
          For the benzene ring the data was interpolated between the times $n\tau$ to better compare the area $t \ApproxEq\tau$.
          In the transition problem $\FDP^\PSI(t)$, $\FDP^\FDA(t)$ and the Zeno approximation agree almost perfectly for $t\gtrapprox \tau$, but differ for small times (see B).
          In the return problem a factor four must be introduced to $\FDP^\PSI(t)$ to match the stroboscopic data (orange triangles).
          This also holds for the complicated dynamics of the random Hamiltonian.
          \label{fig:FDP}
        }
      \end{figure}
      In Fig.~\ref{fig:FDP} we start with plotting the pdfs $\FDP(t)$ of all three models.
      For the random Hamiltonian we took $\tau=0.25\hbar/\gamma$ and for the others the relatively large value $\tau=0.5\hbar/\gamma$.
      We compare the data from the stroboscopic detection protocol with the solution of the non-Hermitian 
      Schr\"odinger equation.
      (For the Benzene Hamiltonian, the stroboscopic data was interpolated as explained in App.~\ref{app:Numerics}.)
      The first column shows the transition problem where the initial and detection states are different.
      We see that both approaches lead to almost the same pdf, except in a boundary layer of size $\Landau{\tau}$ near $t=0$.
      For these small times, the equivalence between both approaches loses its validity.
      The second column shows the data for the return problem.
      Here, the non-Hermitian data is off by a factor of four, just as described in the last section.
      When this correction factor of four is introduced (orange triangles), we again find a nice data collapse,
      except in a small boundary layer around the origin.
      The boundary layer vanishes as $\tau$ goes to zero.
      These observations even hold for the comparably complicated dynamics of the random system and for the infinite line, where the spectrum is continuous.

    \subsection{Normalization and moments}
      \begin{figure}
        \includegraphics[width=0.99\columnwidth]{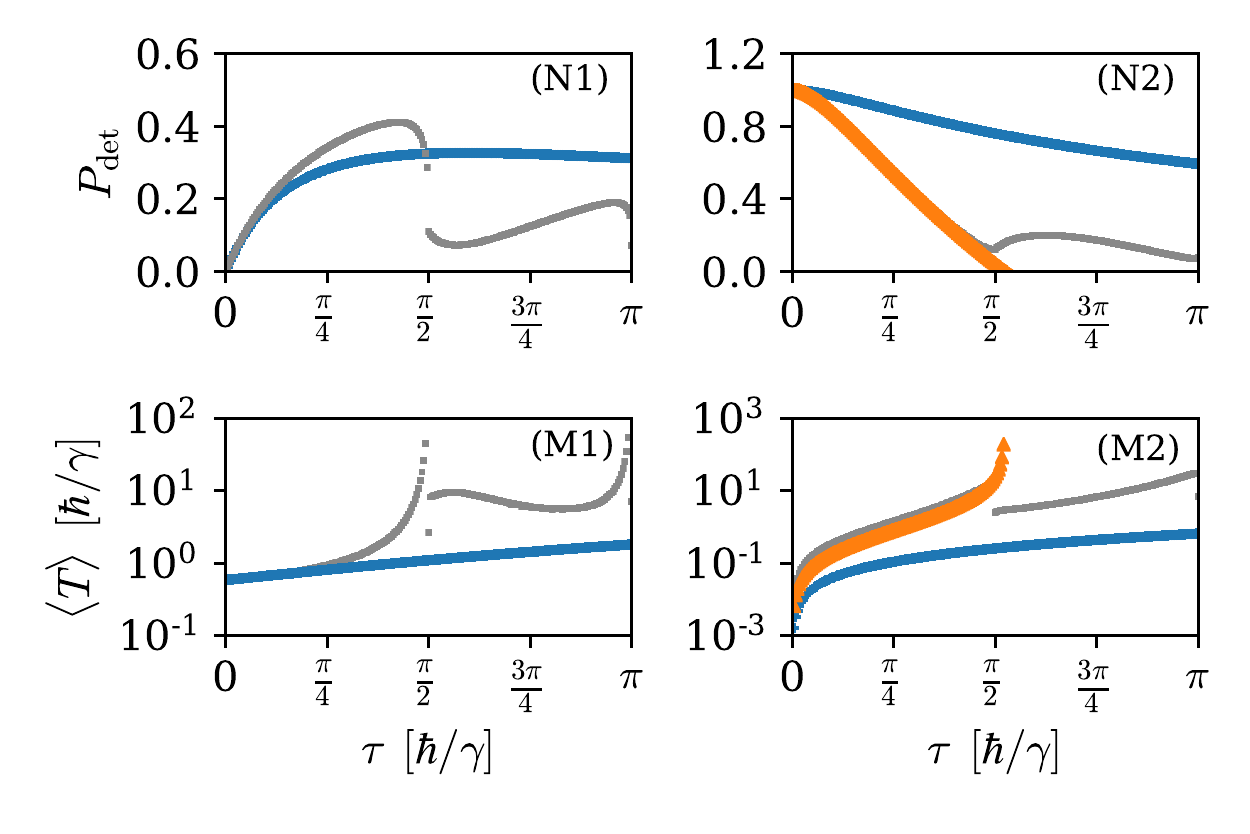}
        \caption{
          Total detection probability (N) and mean first detection time (M) for the infinite line model $\Ham_L$.
          Detection state is $\ket{\PsiDet} = \ket{0}$, initial state is $\ket{1}$ (1) and $\ket{0}$ (2).
          Stroboscopic data (gray squares), non-Hermitian data (blue $+$), and corrected non-Hermitian data (orange $\triangle$, only for return problem).
          Singularities in the stroboscopic data are due to resonant detection periods.
          In the transition problem both approaches coincide for small $\tau$.
          For the return problem, the stroboscopic and corrected data coincide almost perfectly until $\tau = (\pi/2)\hbar/\gamma$, where the corrected total detection probability also becomes negative.
          \label{fig:InfLineStat}
        }
      \end{figure}
      Let us now compare the moments of the distributions $\FDP^\PSI(t)$ and $\FDP^\FDA(t)$.
      In Fig.~\ref{fig:InfLineStat} we plotted the total detection probability (N) and the mean first detection time (M) for the infinite line.
      (Higher moments do not exist, due to the power law decay of the pdf \cite{Thiel2018a}.)
      The left column, Fig.~\ref{fig:InfLineStat}(1), shows data for the initial state $\ket{\PsiIn} = \ket{1}$, where the non-Hermitian and the stroboscopic data clearly only coincide for small $\tau$.
      The right column, Fig.~\ref{fig:InfLineStat}(2), on the other hand, depicts the return problem, when $\ket{\PsiIn} = \ket{\PsiDet} = \ket{0}$.
      Here, the pure non-Hermitian data and the stroboscopic data only share its value at $\tau=0$.
      The corrected result of Eq.~\eqref{eq:TDPMomentEquiv}, however, fits almost perfectly until 
      $\tau_{3/4} \ApproxEq (\pi/2)\hbar/\gamma$.
      This is a pleasant surprise.
      In fact, as we show in the appendix, the difference between $\TDP^\PSI(\PsiDet)$ and $\TDP^\FDA(\PsiDet)$ is $\Landau{\tau^6}$, and so is tiny for small $\tau$.
      It is clear that the two results must diverge from each other beyond $\tau_{3/4}$, where $\TDP^\PSI(\tau_{3/4}) = 3/4$, and thus $4\TDP^\PSI(\tau_{3/4}) - 3 = 0$, i.e. when the corrected expression Eq.~\eqref{eq:TDPMomentEquiv} becomes negative and thus non-physical.

      $\tau_c = (\pi/2) \hbar/\gamma$ is also a special resonant value for the infinite line model \cite{Friedman2017b, Thiel2018a}.
      Starting from this value, the Hamiltonian's spectrum stretched by a factor $\tau$ does not 
      ``fit around the unit circle'' anymore.
      Then the spectrum of $\Ham$ and $\TEO(\tau) = e^{-i\tau \Ham/\hbar}$ start to become fundamentally different,
      because eigenstates of the Hamiltonian around the band edges become dynamically equivalent.
      This aliasing effect can not be mapped to the non-Hermitian system.
      Still, we find an almost perfect data collapse between the stroboscopic data and the corrected non-Hermitian data 
      of Eq.~\eqref{eq:TDPMomentEquiv} for $\tau<\tau_c$ in Fig.~\ref{fig:InfLineStat}.
      
\begin{figure}
        \includegraphics[width=0.99\columnwidth]{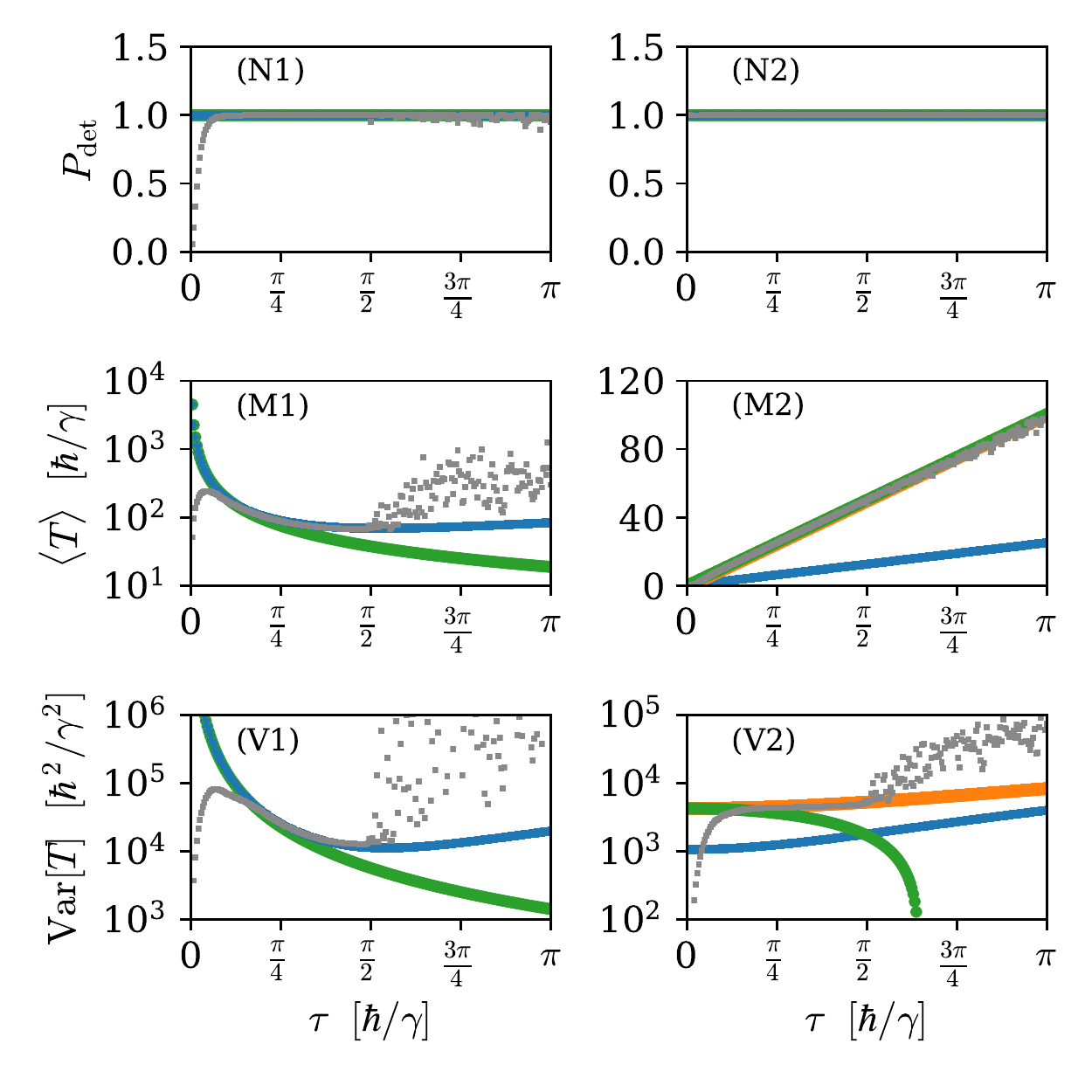}
        \caption{
          First detection statistics for the random Hamiltonian $\Ham_R$.
          Detection state is given by Eq.~\eqref{eq:DefPsiDetEx}.
          Total detection probability (N), mean first detection time (M), and its variance (V),
          for the initial states given by Eq.~\eqref{eq:RandPsiIn}.
          Stroboscopic data (gray squares), non-Hermitian data (blue $+$), corrected non-Hermitian 
          data (orange $\triangle$), and Zeno limit (green $\circ$).
          Fluctuations in the stroboscopic data for $\tau>\pi/2(\hbar/\gamma)$ correspond to correspond to resonant detection times, which can not be mapped to the non-Hermitian picture.
          The dip in the stroboscopic data for $\tau\to0$ is a numerical artifact due to slow convergence (see appendix~\ref{app:Numerics}).
          For small $\tau$ all approaches give the same result.
          All approaches give the correct total detection probability (top row, N), except for the resonant detection periods.
          In the transition problem (left column, 1), the non-Hermitian approach gives good results roughly until the first resonance.
          In the return problem (right column, 2), the corrected non-Hermitian data or the Zeno approximation is more appropriate.
          The latter two describe the mean perfectly for almost all $\tau$ (M2).
          The non-Hermitian approach describes the variance better for intermediate $\tau$ than the Zeno approximation (V2).
          \label{fig:RandStat}
        }
      \end{figure}

\begin{figure*}
        \includegraphics[width=0.99\textwidth]{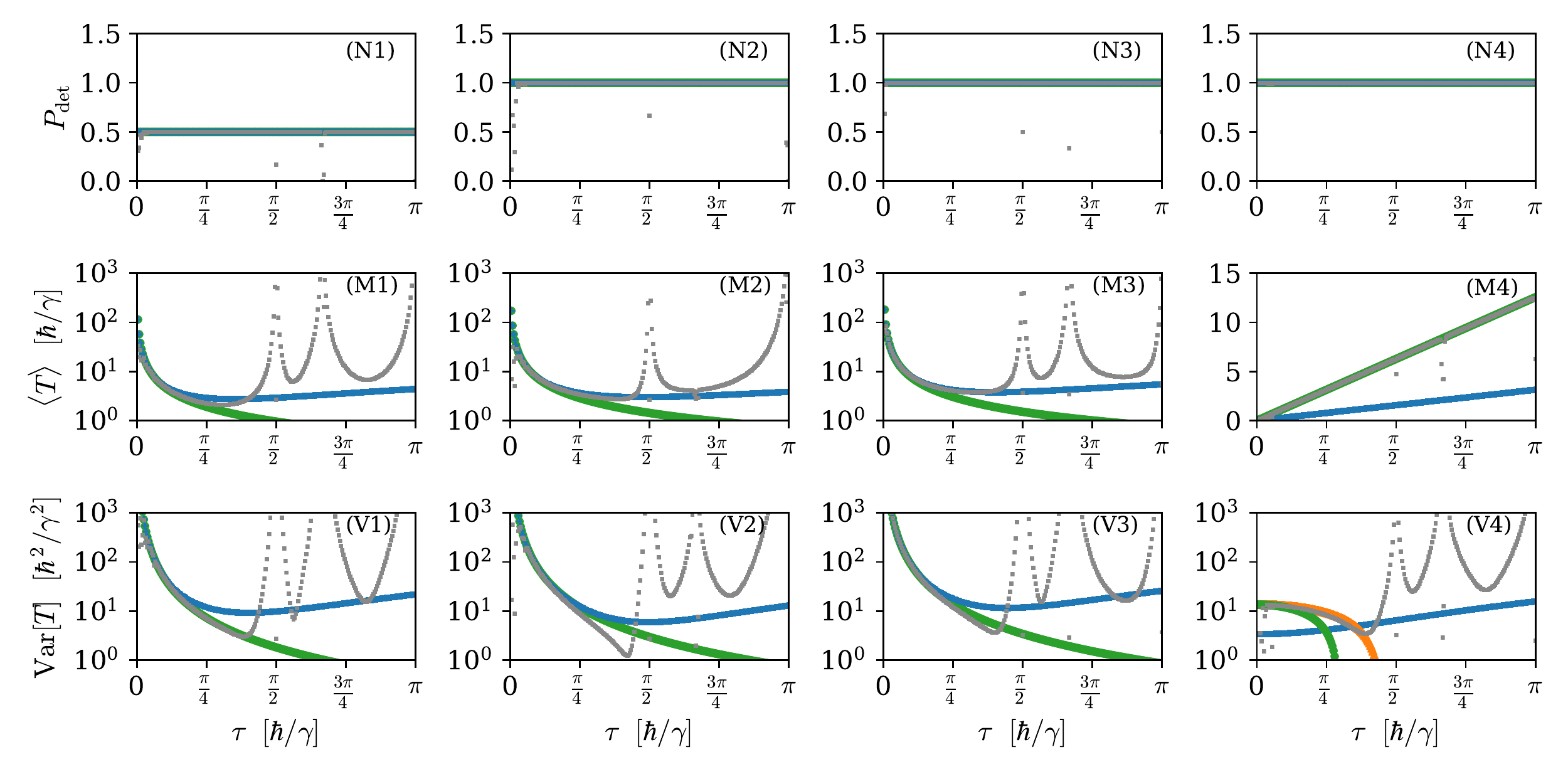}
        \caption{
          First detection statistics for the benzene ring $\Ham_B$, Eq.~\eqref{eq:BenzHam}.
          Detection state is given by Eq.~\eqref{eq:DefPsiDetEx} and initial states are given by Eq.~\eqref{eq:BenzPsiIn}.
          The notation is as in Fig.~\ref{fig:RandStat} and the same conclusions hold here.
          Just like in that figure, the correction becomes necessary in the return problem (right-most column, 4), and describes the stroboscopic data in a larger $\tau$-range than the Zeno approximation (V4).
          The divergences and dips in the stroboscopic data are due to resonant detection periods.
          \label{fig:BenzStat}
        }
      \end{figure*}

      Systems with discrete energy spectra also feature critical detection periods, which are defined by the 
      resonance condition $(E_l - E_{l'})\tau = 0 \mod 2\pi \hbar$, when two energy levels become equivalent in $\TEO(\tau)$.
      In these systems, there is no $\tau$-dependence in $\TDP$ \cite{Thiel2019a}, except at these exceptional values.
      The correction \eqref{eq:TDPMomentEquiv} to $\TDP^\PSI$ becomes meaningless, because the return problem yields $1 = \TDP^\FDA = \TDP^\PSI$,
      which results in $4\TDP^\PSI - 3 = 1$ as well.
      However, the correction factor is still important for the moments, as Figs.~\ref{fig:RandStat} 
      and \ref{fig:BenzStat} show for the random Hamiltonian and the benzene ring, respectively.
      Both figures demonstrate how the non-Hermitian approach correctly predicts the total detection probability except at resonant detection periods, a hardly surprising exception.
      Higher moments of the stroboscopic detection protocol are well described by the non-Hermitian approach
      roughly until the first resonance, when both curves depart from each other.
      By virtue of our normalization of each system's energy spectrum, this first resonance lies at $\tau_c = (\pi/2)\hbar/\gamma$.
      We also see that the non-Hermitian data is not appropriate in the return problem, where the correction Eq.~\eqref{eq:TDPMomentEquiv}
      must be used.
      This is particularly apparent in the return problem's mean first detection time (M), which was shown to be quantized in Ref.~\cite{Gruenbaum2013a}.
      There, we have $\EA{T}^\FDA = w \tau$ (see below), but $\EA{T}^\PSI = w\tau/4$.
      Introducing the correction factor four makes the curves for the mean collapse for almost all $\tau$.

      As we have demonstrated numerically, the non-Hermitian and the stroboscopic approach give equivalent results for small detection periods.
      In the return problem, however, the non-Hermitian data needs to be corrected.
      In general, they only coincide (after correction, in the case of the return problem) up to a relative error of order $\sLandau{\tau^2}$.
      From this point of view, the good numerical correspondence is rather surprising.

  \section{Discrete energy spectra and the Electrostatic Formalism}
   \label{sec:Discrete}
    From the perspective of the equivalence between $\FDP^\FDA(t)$ and $\FDP^\PSI(t)$, one is still left with finding the distribution $\FDP^\PSI(t)$, which poses a considerable problem for a general system.
    More tractable is the question of the actual small $\tau$ limit of both systems.
    This question will be answered in Sec.~\ref{sec:Ultimate} for finite-dimensional systems with a discrete energy spectrum.
    As a first step, we will derive some general features that all finite-dimensional systems have in common.
    This will reveal the special nature of the return problem and help us to take the Zeno limit in the next section.

    In this section, we focus on systems with discrete energy spectra.
    These systems admit the usual diagonal form of the Hamiltonian:
    \begin{equation}
      \Ham
      =
      \Sum{l}{} E_l \hat{P}_l
      =
      \Sum{l}{} E_l \sSum{m=1}{g_l} \dyad{E_{l,m}}
      ,
    \label{eq:HamDiag}
    \end{equation}
    where $E_l$ are the energy levels, each of them $g_l$-fold degenerate with eigenstates $\ket{E_{l,m}}$,
    and eigenspace projector $\hat{P}_l$.
    Since we explicitly account for the degeneracies from the start, all energy levels are distinct $E_l \ne E_{l'}$.

    We first consider the stroboscopic detection protocol and briefly review the results originally reported in Ref.~\cite{Gruenbaum2013a}.
    Later, we apply the same results to the non-Hermitian setup.

    \subsection{The stroboscopic detection protocol}
      \paragraph{Formal solution for the pdf}
        Using the diagonal form, Eq.~\eqref{eq:HamDiag}, in the definitions of Eq.~\eqref{eq:DefResolventStrobo} yields:
        \begin{equation}
          u_\FDA(z)
          = 
          \Sum{l=1}{w} \frac{p_l}{1 - ze^{-i\frac{\tau E_l}{\hbar}}}
          \qc
          v_\FDA(z)
          = 
          \sSum{l=1}{w} \frac{p_l q_l}{1 - z e^{-i\frac{\tau E_l}{\hbar}}}
          .
        \label{eq:DiscResStrobo}
        \end{equation}
        where $p_l := \ev*{\hat{P}_l}{\PsiDet}$ and $p_l q_l := \mel*{\PsiDet}{\hat{P}_l}{\PsiIn}$.
        It should be noted that not all energy levels $E_l$ actually contribute to the problem.
        Any energy level which has no overlap with the detection state, i.e. for which $\hat{P}_l\ket{\PsiDet} = 0$, will appear neither in $u_\FDA(z)$ nor in $v_\FDA(z)$.
        They can safely be ignored.
        $\Ham$ could even possess a continuous part of the spectrum,
        as long as it has no overlap with the detection state.
        This is what we mean when we talk about systems ``with a discrete spectrum''.\footnote{
          Mathematically speaking, our truncation/construction of $\Ham$ ensures that $\ket{\PsiDet}$
          is a cyclic vector of $\Ham$ and that the space of vectors $\{ \ket{\PsiDet}, \Ham \ket{\PsiDet}, \Ham^2 \ket{\PsiDet}, \hdots \}$
          has dimension $w$.
        }
        We assume that $w$ different energy levels appear in $u_\FDA(z)$, and that for each of those $p_l > 0$.
        Naturally, we have $\sSum{l=1}{w} p_l = \ip{\PsiDet} = 1$, so that none of them can be larger than unity.
        A similar normalization holds for the $q_l$, namely $\sSum{l=1}{w} p_l q_l = \ip*{\PsiDet}{\PsiIn}$.
        Apart from that, the $q_l$ can be arbitrary complex numbers.
        Another implicit assumption in Eq.~\eqref{eq:DiscResStrobo} is that resonant $\tau$ are avoided, such that 
        all phase factors $e^{-i\tau E_l/\hbar}$ are unique and no pair of terms yields the same denominator.
        Otherwise, $w$ would need to be redefined.
        
        The points $z=e^{i\tau E_l/\hbar}$ are simple poles of the functions $u_\FDA(z)$ and $v_\FDA(z)$.
        Nevertheless, these poles cancel in $\FDA(z)$, which is analytic in the unit disk.
        Still $\FDA(z)$ has $w-1$ poles $\mathfrak{z}_l$ outside the unit disk defined by 
        \begin{equation}
          0 = u_\FDA(\mathfrak{z}_l)
          .
        \label{eq:DefPolesStrobo}
        \end{equation}
        These simple poles are  crucial to writing down a formal solution to $\FDA_n$.
        This is achieved by a partial fraction decomposition of $\FDA(z)$.
        Note that when some rational function $h(z) = f(z)/g(z)$ has the simple poles $\mathfrak{z}_l$ (which are simple zeros of $g(z)$), 
        then it admits the decomposition $h(z) = \sSum{l}{} \frac{C_l}{z-\mathfrak{z}_l}$, and the coefficients 
        can be obtained via Heaviside's formula:
        \begin{equation}
          C_l = \lim_{z\to \mathfrak{z}_l} (z-\mathfrak{z}_l) h(z) = \frac{f(\mathfrak{z}_l)}{g'(\mathfrak{z}_l)}
          .
        \label{eq:Heaviside}
        \end{equation}
        Applying this formula to $\FDA(z)/z$ with Eq.~\eqref{eq:DefGenFunc} 
        [so that $g(z) = u_\FDA(z)$ and $f(z) = (v_\FDA(z)-\ip{\PsiDet}{\PsiIn})/z$] gives:
        \begin{equation}
          \FDA(z)
          =
          - \Sum{l=1}{w-1} 
          \frac{v_\FDA(\mathfrak{z}_l) - \ip{\PsiDet}{\PsiIn}}{ \mathfrak{z}_l u_\FDA'(\mathfrak{z}_l)}
          \frac{\tfrac{z}{\mathfrak{z}_l}}{1 - \tfrac{z}{\mathfrak{z}_l}}
          .
        \label{eq:GenFuncDisc}
        \end{equation}
        Expanding each geometric series gives $\FDA_n$:
        \begin{equation}
          \FDA_n
          =
          -
          \Sum{l=1}{w-1} 
          \frac{
            v_\FDA(\mathfrak{z}_l) - \ip{\PsiDet}{\PsiIn}
          }{
            u_\FDA'(\mathfrak{z}_l)
          }
          \mathfrak{z}_l^{-n-1}
          .
        \label{eq:DiscreteAmpStrobo}
        \end{equation}
        Hence, knowledge about the poles $\mathfrak{z}_l$ of $\FDA(z)$ results in a decomposition of 
        the detection amplitudes in terms of exponentially decaying modes.
        (They decay rather than grow because $\abs{\mathfrak{z}_l}>1$.)

      \paragraph{Electrostatic analogy}
        The poles themselves can be found from a nice electrostatic analogy.
        The starting point is Eq.~\eqref{eq:DefPolesStrobo}, which has a trivial solution $\mathfrak{z} = \infty$.
        When this is removed by multiplication by $\mathfrak{z}$, one arrives at:
        \begin{equation}
          0
          =
          \mathfrak{z} u_\FDA(\mathfrak{z})
          =
          \Sum{l=1}{w} \frac{p_l}{\frac{1}{\mathfrak{z}} - e^{-i\frac{\tau E_l}{\hbar}}}
          .
        \label{eq:PreElectro}
        \end{equation}
        We now construct the 2D-electrostatic potential 
        \begin{equation}
          V_\FDA(x,y)
          :=
          \Sum{l=1}{w} p_l \ln \sqrt{ [x - \cos(\tfrac{\tau E_l}{\hbar})]^2  + [y+\sin(\tfrac{\tau E_l}{\hbar})]^2 }
        \label{eq:DefElectroStrobo}
        \end{equation}
        by placing 2D-point charges of magnitude $p_l$ on each eigenvalue eigenvalue $e^{-i\tau E_l/\hbar}$ of $\TEO(\tau)$ on the unit circle.
        (Here, we use the canonical mapping between the real and the complex plane: $\Reals^2 \ni (x,y) \leftrightarrow x+iy \in \Complexs$.)
        Then the points $\mathfrak{z}_V = 1/\mathfrak{z} = x + iy$, are seen to be $V_\FDA(x,y)$'s stationary points, i.e. the points of vanishing gradient $\Nabla_{(x,y)}V_\FDA(x,y) = \V{0}$.
        These gradient equations are exactly the real and imaginary parts of Eq.~\eqref{eq:PreElectro}.
        
        In the return problem, we have $v_\FDA(z)= u_\FDA(z)$ and there is a relation between the poles $\mathfrak{z}$ 
        and the zeros $\mathfrak{n}$ of $\FDA(z)$, because there is a symmetry between the $u_\FDA(z)$ and its conjugate $u_\FDA^*(z)$:
        \begin{equation}
          u_\FDA(z)
          =
          -
          \qty[ u_\FDA^*\qty(\tfrac{1}{z}) - 1 ]
          .
        \label{eq:ZeroPoleRelation}
        \end{equation}
        (Remember that $u_\FDA^*(z) = [u_\FDA(z^*)]^*$.)
        Eq.~\eqref{eq:ZeroPoleRelation} is easily seen from the definition of $u_\FDA(z)$ and the unitarity of $\TEO(\tau)$.
        So if $\mathfrak{z}$ is a root of $u_\FDA(\mathfrak{z}) = 0$ (and thus a pole of $\FDA(z)$, then $\mathfrak{n} = 1/\mathfrak{z}^*$ is a root of $u_\FDA(\mathfrak{n}) - 1= 0$ [and thus a zero of $\FDA(z)$ for the return problem, see Eq.~\eqref{eq:DefGenFunc}].
        These zeros $\mathfrak{n} = \mathfrak{z}_V^*$ are the conjugated stationary points of the two-dimensional electrostatic potential $V_\FDA(x,y)$.
        The operation $\mathfrak{n} = 1/\mathfrak{z}^*$ that connects the zeros and poles is a reflection about the unit circle.
        Note, that $\FDA(z)$ has an additional trivial zero at $z=0$, because $u_\FDA(0) = 1$, which is not mapped by the electrostatic analogy.
        Furthermore, note that $\FDA(z)$ for the transition problem has different zeros than these stationary points.
        These, however, are not as important to the first detection statistics, because only the poles $\mathfrak{z}$ determine the decay modes.

      \paragraph{The return problem}
        In the return problem, when $\ket{\PsiIn} = \ket{\PsiDet}$ and $v_\FDA(z) = u_\FDA(z)$, the knowledge of the poles is sufficient to describe {\em all} first detection statistics.
        The electrostatic potential can then be used to describe these statistics \cite{Yin2019a, Liu2020a}.
        The symmetry relation \eqref{eq:ZeroPoleRelation} is a peculiarity of the return problem
        and implies that $\FDA^*(1/z) = 1/\FDA(z)$, when $\ket{\PsiIn} = \ket{\PsiDet}$.
        It follows from Eq.~\eqref{eq:TDPStrobo} that the return state is almost surely detectable:
        \begin{align}
          \TDP(\PsiDet)
          = &
          \ointctrclockwise\limits_\Cauchy
          \frac{\dd z}{2 \pi i z}
          \frac{\FDA(z)}{\FDA(z)}
          = 1
          .
        \label{eq:SureTDPStrobo}
        \end{align}
        With the same identity, we find that the mean first detection time is a contour integral over a logarithmic derivative:
        \begin{align}
          \EA{T}^\FDA
          = &
          \tau
          \ointctrclockwise\limits_\Cauchy
          \frac{\dd z}{2 \pi i}
          \frac{\dd }{\dd z} 
          \ln \FDA(z)
          = w \tau
          .
        \label{eq:QuantMeanStrobo}
        \end{align}
        By virtue of the argument principle of complex analysis, the contour integral over a logarithmic derivative is equal to the number of the function's zeros minus the number of poles.
        This number is $w$.
        $\FDA(z)$ has no poles inside the unit disk.
        There are $w-1$ zeros found from the stationary points of Eq.~\eqref{eq:DefElectroStrobo} and one trivial zero at $z=0$,
        where $u_\FDA(z=0) = 1$.
        Therefore the mean first detection time of the return problem in the stroboscopic detection protocol is quantized and equal to the number of energy levels that appear in $\ket{\PsiDet}$ \cite{Gruenbaum2013a, Sinkovicz2015a, Friedman2017a}.
        Eqs.~(\ref{eq:SureTDPStrobo}, \ref{eq:QuantMeanStrobo}) are nicely demonstrated in 
        Figs.~\ref{fig:RandStat}(2) and \ref{fig:BenzStat}(4), where the total detection probability 
        and the mean show constant and linear behavior, respectively.

      Finding the zeros of $u_\FDA(z)$ or the stationary points of $V_\FDA(x,y)$ is in general
      equally hard and not possible outside of some perturbative limit.
      Nevertheless, the above arguments -- which are known for the stroboscopic detection protocol \cite{Gruenbaum2013a} -- can be applied directly to the non-Hermitian Schr\"odinger equation.
      There we can gain new insights.

    \subsection{The non-Hermitian Schr\"odinger equation}
      \paragraph{Formal solution for the pdf}
        Again, we express the resolvents in terms of the overlaps $p_l$, and $q_l$ as well as 
        with the energy levels $E_l$:
        \begin{equation}
          u_\PSI(s)
          =
          \Sum{l=1}{w} \frac{p_l}{s + i \frac{E_l}{\hbar}}
          \qc 
          v_\PSI(s)
          =
          \sSum{l=1}{w} \frac{p_l q_l}{s + i \frac{E_l}{\hbar}}
        \label{eq:DiscResSchroedi}
        \end{equation}
        As before, $v_\PSI(s)$ and $u_\PSI(s)$ have poles at $-i E_l / \hbar$, which cancel in $\PSI(s)$.
        $\PSI(s)$ has $w$ simple poles $\mathfrak{s}_l$ in the left half plane defined by the equation:
        \begin{equation}
          0 = 1 + \frac{2}{\tau} u_\PSI(\mathfrak{s}_l)
        \label{eq:DefPolesSchroedi}
        \end{equation}
        Using these poles, we can write down a partial fraction decomposition of $\PSI(s)$:
        \begin{equation}
          \PSI(s)
          =
          \frac{\tau}{2}
          \Sum{l=1}{w} 
          \frac{v_\PSI(\mathfrak{s}_l) }{u_\PSI'(\mathfrak{s}_l)}
          \frac{1}{s - \mathfrak{s}_l}
          .
        \label{eq:}
        \end{equation}
        Using the residue theorem, the inverse Laplace transform is easily performed:
        \begin{equation}
          \PSI(t)
          =
          \frac{\tau}{2}
          \Sum{l=1}{w}
          \frac{v_\PSI(\mathfrak{s}_l) }{u_\PSI'(\mathfrak{s}_l)}
          e^{t \mathfrak{s}_l}
          .
        \label{eq:DiscreteAmpSchroedi}
        \end{equation}
        Since all poles lie in the left-half plane, such that $\Re[\mathfrak{s}_l] < 0$, all exponentials are decaying.
        Note that Eq.~\eqref{eq:DefPolesSchroedi} has $w$ roots as opposed to Eq.~\eqref{eq:DefPolesStrobo}.

      \paragraph{Electrostatic analogy}
        Again, we can find an electrostatic analogy by inspecting Eq.~\eqref{eq:DefPolesSchroedi}.
        This time, the conjugated poles $\mathfrak{s}^* = x + iy$ are given by the stationary points of the following electrostatic potential:
        \begin{equation}
          V_\PSI(x,y)
          :=
          \frac{\tau}{2} x
          +
          \Sum{l=1}{w}
          p_l
          \ln\sqrt{ x^2 + \qty( y - \tfrac{E_l}{\hbar} )^2 }
          .
        \label{eq:DefElectroSchroedi}
        \end{equation}
        Here a positive point charge of magnitude $p_l$ is placed on each eigenvalue of the {\em Hamiltonian} on the imaginary axis.
        In contrast to the previous potential, there is an additional constant force proportional to $\tau$.

\begin{figure}
          \includegraphics[width=0.9\columnwidth]{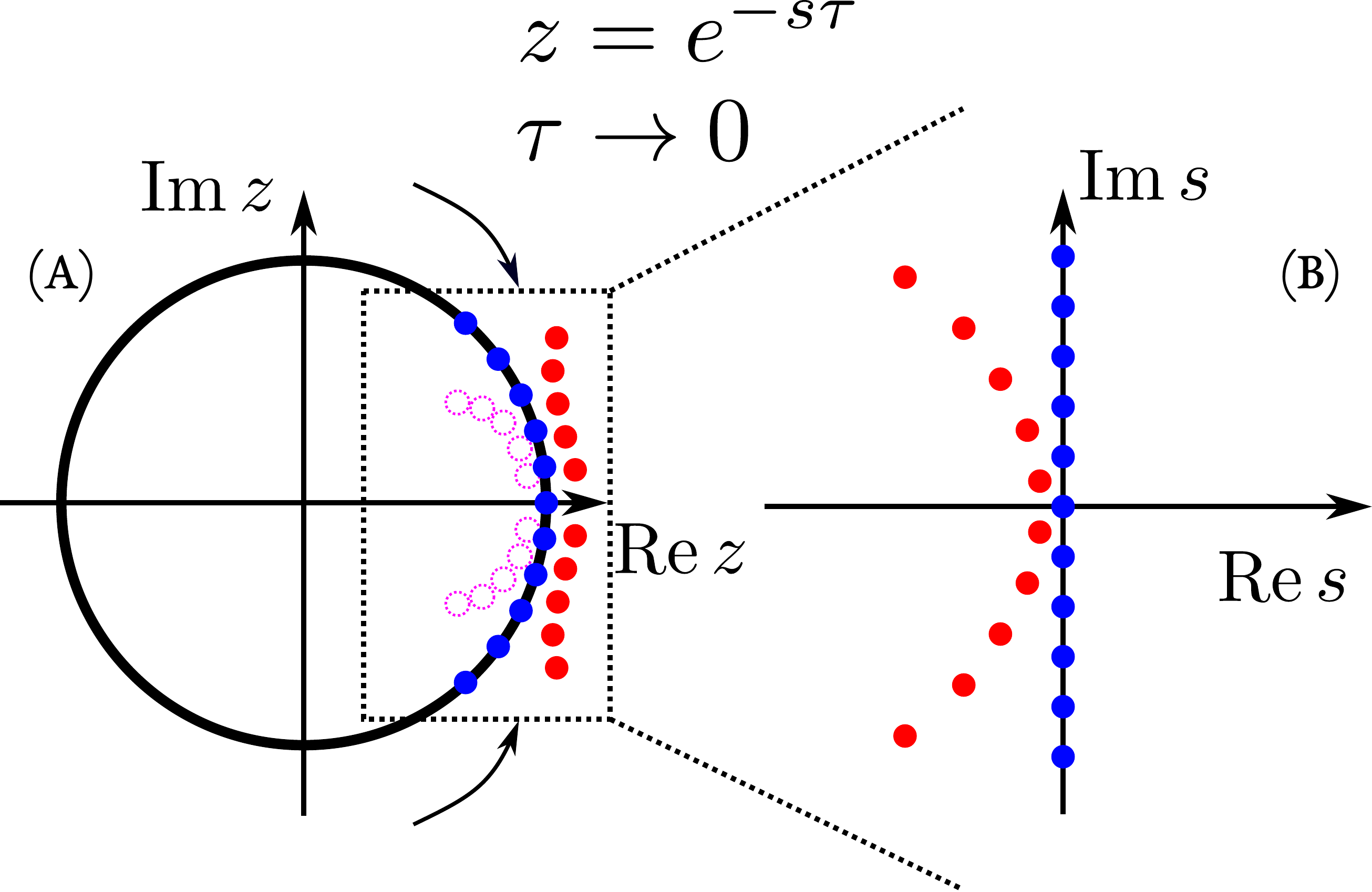}
          \caption{
            Poles in the electrostatic analogy.
            (A) The eigenvalues $e^{-iE_l \tau/\hbar}$ of $\TEO(\tau)$ lie on the unit circle (blue).
            The poles $\mathfrak{z}_l$ of $\FDA(z)$ (red) lie outside the unit circle.
            Their mirror images $\mathfrak{n}_l$ (empty circles) lie inside the unit disk.
            After equipping every one of $\TEO(\tau)$'s eigenvalues with an electric charge
            equal to the overlap $p_l = \ev*{\hat{P}_l}{\PsiDet}$, Ref.~\cite{Gruenbaum2013a} finds the mirrored poles 
            as the stationary points of the electrostatic potential \eqref{eq:DefElectroStrobo}.
            As $\tau$ decreases all points move to $z = 1$ (arrows).
            (B) Our mapping $z=e^{-s\tau}$ ``zooms in'' around $z=1$.
            The unit circle's curvature is replaced by a constant force, see Eq.~\eqref{eq:DefElectroSchroedi}.
            The charges are placed at $iE_l/\hbar$ on the imaginary axes.
            The poles $\mathfrak{s}_l$ are the complex conjugates of the new electrostatic potential's stationary points.
            \label{fig:PoleMapping}
          }
        \end{figure}

        Remember from sec.~\ref{sec:Equivalence} that we related $z$ and $s$ via $z = e^{-s\tau}$ and a subsequent 
        small $\tau$ expansion.
        The exponential maps the outside of the unit circle to the left half of a strip $\{ s \in \Complexs | \Re[s] <0\qc \abs{\Im[s]} < \pi/\tau \}$.
        Taking $\tau\to0$ enlarges this domain to the complete left half plane.
        During this procedure, the charges, which were originally on the unit circle, move to the imaginary axis.
        The curvature of the unit circle, which forced the zeros inside the unit disk, gets mapped to the constant-force term,
        which forces the poles into the left-half plane.
        This ensures the boundedness of $\PSI(t)$.
        See also Fig.~\ref{fig:PoleMapping}.
        Another way to relate Eqs.~\eqref{eq:DefPolesStrobo} and \eqref{eq:DefPolesSchroedi} is via Eq.~\eqref{eq:ResCorrespondence},
        which reveals the latter as a straightforward small-$\tau$ version of the former.
        The advantage of $V_\PSI(x,y)$ over its counterpart $V_\FDA(x,y)$ is the much easier geometry.
        All the charges lie on a line.
        This makes it possible to find all the poles in the Zeno limit, as will be presented in the next section.

        Similar to the stroboscopic case, there is a relation between $u_\PSI(s)$ and its conjugate:
        \begin{equation}
          - u_\PSI^*(-s) = u_\PSI(s)
        \label{eq:ConjugateRelationSchroedi}
        \end{equation}
        Using this equation one can relate the poles of $\PSI(s)$ with those of $\PSI^*(-s)$ in Eqs.~\eqref{eq:DefTDPSchroedi} and \eqref{eq:DefMomentSchroedi}.

        Just as in the stroboscopic case, one finds a quantized mean first detection time 
        \begin{equation}
          \EA{T}^\PSI = \frac{w}{4} \tau
        \label{eq:}
        \end{equation}
        from complex function arguments.
        This is -- as far as we know -- a new result and will be explored in more depth in another publication \cite{Thiel2019f}.
        We did not find an equally nice analytical demonstration that $\TDP^\PSI(\PsiDet) = 1$ for the non-Hermitian setup.
        This is however evident from our numerics in Figs.~\ref{fig:RandStat}(N2) and \ref{fig:BenzStat}(N4), 
        and will be shown in the limit $\tau\to0$ in the next section.
        We also demonstrate it in the limit $\tau\to\infty$ in App.~\ref{app:Lazy}.

  \section{The Zeno limit in the Electrostatic Formalism}
  \label{sec:Ultimate}
    \subsection{Non-Hermitian Schr\"odinger equation}
\begin{figure}
        \includegraphics[width=0.99\columnwidth]{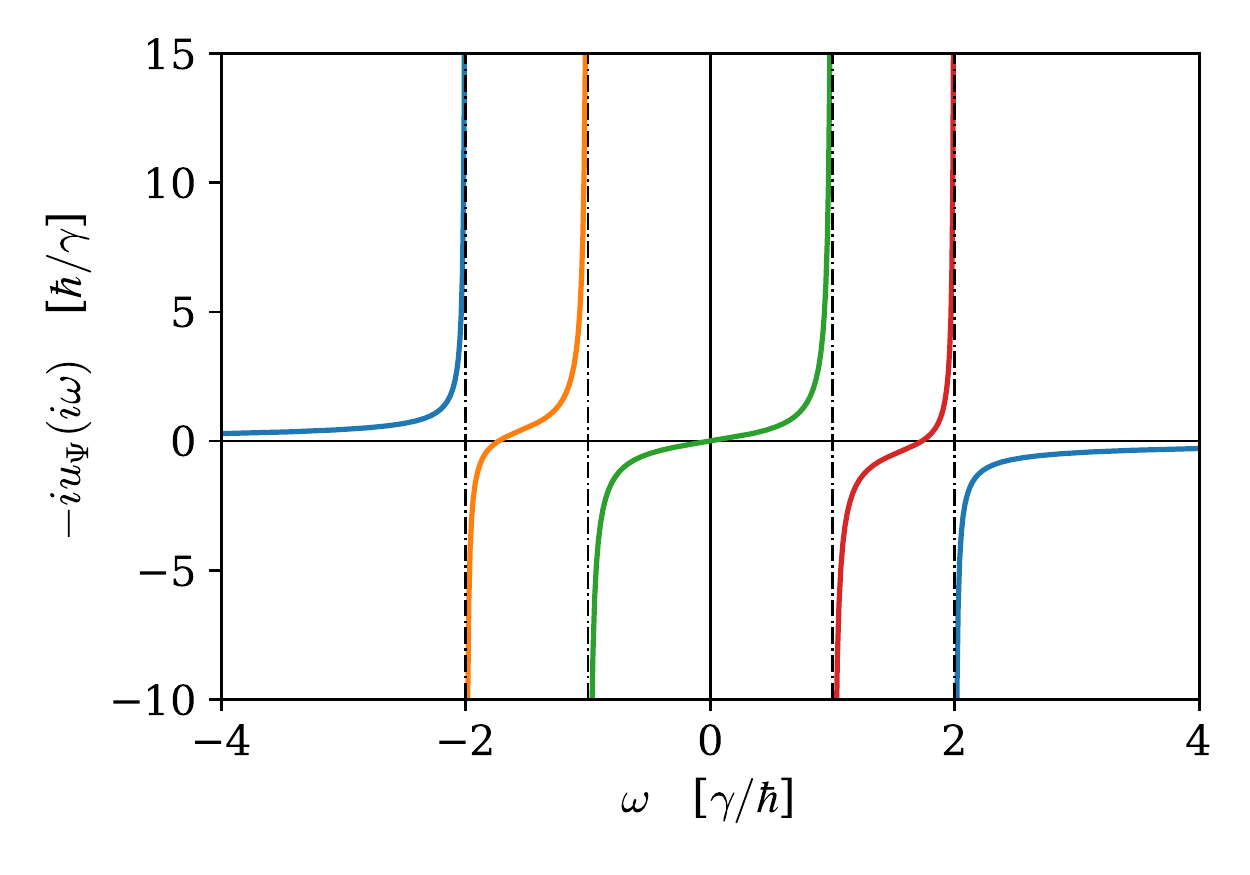}
        \caption{
          The resolvent $u_\PSI(s)$ for the benzene ring $\Ham_B$.
          We show $\omega \mapsto -iu_\PSI(i\omega)$.
          The dash-dotted lines, where $u_\PSI(s)$ diverges show the negative energy levels.
          The zeros of this function are the negative absorption frequencies $-\omega_l$.
          $u_\PSI(s)$'s derivative at these points is related to the absorption rates $\lambda_l\tau$.
          Clearly, one zero is located between two adjacent energy levels, and the function is monotonic in this interval as well.
          Therefore, the absorption frequencies are easily found numerically.
          \label{fig:Contour}
        }
      \end{figure}

      Starting with the electrostatic analogy, we can find all the poles when $\tau$ is sufficiently small.
      Consider $V_\PSI(x,y)$ from Eq.~\eqref{eq:DefElectroSchroedi} first for vanishing $\tau$.
      Since all charges have the same sign, and there is no constant force, the stationary points must lie on the imaginary axis, i.e. $x=0$, one between each pair of adjacent energies.
      Assume that the $w$ energy levels are ordered, i.e. $E_l < E_{l+1}$.
      We will then find one stationary point of $V_\PSI(x,y)$ 
      at $0+i\omega_l$ with $E_l < \hbar\omega_l < E_{l+1}$.
      We call the $\omega_l$s the absorption frequencies.
      They are found by solving the equation:
      \begin{equation}
        0 
        = 
        - \frac{i}{\hbar} u_\PSI(-i\omega_l) 
        = 
        \ev{\frac{1}{\hbar \omega_l - \Ham}}{\PsiDet}
        ,
      \label{eq:DefOmega}
      \end{equation}
      i.e. they are the zeros of the resolvent.
      We plotted the resolvent for the benzene ring in Fig.~\ref{fig:Contour},
      where it is apparent that the resolvent is a monotonic function between two energy values.
      Therefore, although the absorption frequencies are only defined implicitly,
      they are easy to find in practice.
      They interlace with the energy levels, are thus bracketed, and the defining function is monotonic.
      Any numerical root finding algorithm will find them with ease.
      There are $w-1$ solutions $\omega_l$, $l=1,\hdots,w-1$ to Eq.~\eqref{eq:DefOmega}.
      A similar interlacing for absorption and relaxation rates has been found for the first passage problem of classical random walks \cite{Hartich2019a}.

      Now we consider $\tau>0$ but small.
      The additional small constant force will shift the stationary points from $i\omega_l$ slightly into the left half plane.
      Thus we make the following ansatz for the pole  
      \begin{equation}
        \mathfrak{s}_l \AsymEq - \lambda_l \tau - i\omega_l
      \label{eq:PoleAssumpZenoSchroedi}
      \end{equation}
      and plug it into Eq.~\eqref{eq:DefPolesSchroedi} together with Eq.~\eqref{eq:DiscResSchroedi}.
      This yields:
      \begin{align}
        0
        = & 
        \frac{\tau}{2} 
        +
        \Sum{l'=1}{w}
        \frac{p_{l'}}{ - \lambda_l \tau + \frac{i}{\hbar} ( E_{l'} - \hbar \omega_l ) }
        \\ = & \nonumber
        \frac{\tau}{2} 
        -
        \Sum{l'=1}{w}
        \frac{p_{l'} [\lambda_l \tau + \frac{i}{\hbar} ( E_{l'} - \hbar \omega_l )]}{ \lambda_l^2 \tau^2 + \frac{1}{\hbar^2} ( E_{l'} - \hbar\omega_l )^2 }
        \\ \AsymEq & \nonumber
        \frac{\tau}{2} 
        -
        \Sum{l'=1}{w}
        \frac{p_{l'} [\lambda_l \tau + \frac{i}{\hbar} ( E_{l'} - \hbar \omega_l )]}{\frac{1}{\hbar^2} ( E_{l'} - \hbar\omega_l )^2 }
        \\ = & \nonumber
        \tau \qty{
          \frac{1}{2}
          -
          \hbar^2 \lambda_l \Sum{l'=1}{w} \frac{p_{l'}}{(E_{l'} - \hbar \omega_l)^2}
        }
        -
        i \hbar 
        \Sum{l'=1}{w}
        \frac{p_{l'}}{E_{l'} - \hbar\omega_l}
        .
      \label{eq:}
      \end{align}
      Higher order terms in $\tau$ were neglected to obtain the third line.
      Equating the imaginary part of the last line with zero results in Eq.~\eqref{eq:DefOmega}, which shows that 
      the imaginary part was correctly chosen.
      The absorption rate $\lambda_l$ is determined from the last line's real part:
      \begin{equation}
        \lambda_l
        =
        \qty[
          2 \hbar^2 \Sum{l'=1}{w} \frac{p_{l'}}{(E_{l'} - \hbar \omega_l)^2}
        ]^{-1}
        =
        \frac{1}{2 u_\PSI'(-i\omega_l)}
        .
      \label{eq:DefLambda}
      \end{equation}
      This gives the missing part to Eq.~\eqref{eq:PoleAssumpZenoSchroedi} that determines the poles $\mathfrak{s}_l$ for $l=1,\hdots,w-1$.
      In addition to the poles, we need $u_\PSI'(\mathfrak{s})$ and $v_\PSI(\mathfrak{s})$ for Eq.~\eqref{eq:DiscreteAmpSchroedi}.
      The first is determined by $\lambda_l$ to leading order:
      \begin{equation}
        u_\PSI'(\mathfrak{s}_l)
        \AsymEq
        u_\PSI'(-i\omega_l)
        =
        \frac{1}{2\lambda_l}
        \qc l=1,\hdots,w-1
        .
      \label{eq:LambdaDeriv}
      \end{equation}
      Furthermore, we find $v_\PSI(\mathfrak{s}_l)$ in leading order.
      For the return problem, this is equal to $v_\PSI(\mathfrak{s}_l) = u_\PSI(\mathfrak{s}_l) = -\tau/2$.
      In the transition problem, we obtain:
      \begin{equation}
        v_\PSI(\mathfrak{s}_l)
        \AsymEq
        \mel*{\PsiDet}{\frac{1}{ \frac{i}{\hbar} \Ham - i \omega_l}}{\PsiIn}
        =:
        - i \theta_l
        ,
      \label{eq:LambdaN}
      \end{equation}
      which defines the ``transition times'' $\theta_l$.

      We have thus found $w-1$ poles close to the imaginary axis.
      These determine the slow dynamics in $\PSI(t)$.
      However, Eq.~\eqref{eq:DefPolesSchroedi} admits another pole far in the left half plane,
      which describes the fast dynamics of $\PSI(t)$.
      This pole is given in leading order by:
      \begin{equation}
        \mathfrak{s}_0
        \AsymEq
        - 2/\tau - i \omega_0
        ,
      \label{eq:SpuriousPole}
      \end{equation}
      where $\hbar \omega_0 = \ev*{\Ham}{\PsiDet}$ is the mean energy of the detection state.
      For the derivative $u_\PSI'(\mathfrak{s}_0)$ and for $v_\PSI(\mathfrak{s}_0)$ we find:
      \begin{equation}
        u_\PSI'(\mathfrak{s}_0) \AsymEq - \frac{\tau^2}{4}
        \qc 
        v_\PSI(\mathfrak{s}_0) \AsymEq - \frac{\tau}{2} \ip{\PsiDet}{\PsiIn}
        .
      \label{eq:SpuriousDeriv}
      \end{equation}

      Now we have gathered all the necessary ingredients to write down $\PSI(t)$.
      Using all the just derived results in Eq.~\eqref{eq:DiscreteAmpSchroedi}, we arrive at:
      \begin{equation}
        \PSI(t)
        \AsymEq
        \left\{ \begin{aligned}
          \ip{\PsiDet}{\PsiIn} e^{-t( \frac{2}{\tau} + i\omega_0 )}
          - i \tau
          \Sum{l=1}{w-1}
          \lambda_l
          \theta_l
          e^{-t( \lambda_l \tau + i \omega_l)}
          \\
          e^{-t( \frac{2}{\tau} + i\omega_0 )}
          - \frac{\tau^2}{2}
          \Sum{l=1}{w-1}
          \lambda_l
          e^{-t( \lambda_l \tau + i \omega_l)}
        \end{aligned} \right.
        .
      \label{eq:}
      \end{equation}
      Here the first line corresponds to the transition problem and the second line corresponds to the return problem.
      Note, that the fast dynamics is the same in both situations.
      The slow dynamics, however, is of different order in $\tau$.

      Squaring the wave function and multiplying it with $4/\tau$ gives the pdf $\FDP^\PSI(t)$.
      We have to leading order in $\tau$:
      \begin{equation}
        \FDP^\PSI(t)
        \AsymEq 
        \frac{4}{\tau} \abs{\ip{\PsiDet}{\PsiIn}}^2 e^{- \frac{4t}{\tau} }
        + 
        \left\{ \begin{aligned}
          4 \tau \abs{ \Sum{l=1}{w-1} \lambda_l \theta_l e^{-t(\lambda_l \tau + i\omega_l)} }^2
          \\
          \tau^3 \abs{ \Sum{l=1}{w-1} \lambda_l e^{-t(\lambda_l \tau + i\omega_l)} }^2
        \end{aligned} \right.
        ,
      \label{eq:FDPZenoSchroedi}
      \end{equation}
      where the first line holds for the transition problem and the second line for the return problem.

      Integration of the pdf gives the following values for $\TDP^\PSI$ and the moments $\EA{T}^\PSI$ in the Zeno limit:
      \begin{align}
      \label{eq:TDPZenoSchroedi}
        \TDP^\PSI
        \AsymEq &
        \left\{ \begin{aligned}
          \abs{\ip{\PsiDet}{\PsiIn}}^2 + \Sum{l=1}{w-1} 2 \lambda_l \abs{\theta_l}^2 
          \qc &  \ket{\PsiIn} \ne \ket{\PsiDet} \\
          1
          \qc &  \ket{\PsiIn} = \ket{\PsiDet} \\
        \end{aligned} \right.
        \\
        \EA{T^m}^\PSI
        \AsymEq &
        \left\{ \begin{aligned}
          \frac{m!}{\TDP^\PSI} \Sum{l=1}{w-1} 
          \frac{2\lambda_l \abs{\theta_l}^2 }{( 2 \lambda_l \tau)^m}
          \qc & \ket{\PsiIn} \ne \ket{\PsiDet} \\
          \delta_{m,1}\frac{\tau}{4} +
          \frac{m!}{4\tau^{m-2}} 
          \Sum{l=1}{w-1}
          \frac{2\lambda_l}{(2\lambda_l)^m}
          \qc & \ket{\PsiIn} = \ket{\PsiDet} \\
        \end{aligned} \right.
      \label{eq:MomentZenoSchroedi}
      \end{align}
      The first measurement term is significant only for the first moment, otherwise it is negligible.
      Our Zeno limit reproduces the general results for the return problem, namely $\TDP^\PSI(\PsiDet) = 1$ and $\EA{T}^\PSI = w\tau/4$.

    \subsection{Stroboscopic detection protocol}
       Given the connections found above between the NHH and stochastic protocol, we can read off the solution of the Zeno limit of the stochastic protocol.
      Of course, one can also approach the problem directly.
      To do this, it is necessary, as we did above in Sec.~\ref{sec:MagicFactor}, to treat the first detection attempt separately from the others.
      The other crucial step is to identify the poles $\mathfrak{z}_l$ with the poles $\mathfrak{s}_l$ of the non-Hermitian approach.
      The poles are determined by Eq.~\eqref{eq:DefPolesStrobo}, but we will follow Sec.~\ref{sec:Equivalence} and write $\mathfrak{z} = e^{-\mathfrak{s}\tau}$.
      Expanding for small $\tau$ allows us to use Eq.~\eqref{eq:ResCorrespondence}, so that $0 = 2u_\FDA(e^{-\mathfrak{s}\tau}) \AsymEq 1 + 2 u_\PSI(\mathfrak{s})/\tau$.
      This means that we can use the poles $\mathfrak{s}_l$ from Eq.~\eqref{eq:DefPolesSchroedi}, that we determined explicitly before, via
      \begin{equation}
        \mathfrak{z}_l 
        \AsymEq
        e^{-\mathfrak{s}_l\tau}
        \AsymEq
        e^{\lambda_l\tau^2 + i \omega_l\tau}
        ,
      \label{eq:PoleAssumpZenoStrobo}
      \end{equation}
      for $l=1,\hdots, w-1$.
      As mentioned before, the poles $\mathfrak{s}_l$ and $\mathfrak{z}_l$ are related by our approximation procedure.
      A sketch can be found in Fig.~\ref{fig:PoleMapping}.

      The final pitfall we have to avoid is that $0=u_\FDA(z)$ has $w-1$ solutions, but $0 = 1 + 2u_\PSI(s)/\tau$ has $w$ solutions.
      The non-Hermitian approach has one additional pole, namely the fast mode $\mathfrak{s}_0$, that does not appear in the 
      stroboscopic setup.
      The first measurement and $\FDA_1$ play the role of the fast mode here.
      By simply excluding this spurious pole, and using Eqs.~(\ref{eq:ResCorrespondence}, \ref{eq:NCorrespondence}, 
      \ref{eq:DefPolesSchroedi}, \ref{eq:PoleAssumpZenoSchroedi}, \ref{eq:LambdaDeriv}, \ref{eq:LambdaN}) 
      and \eqref{eq:PoleAssumpZenoStrobo} in Eq.~\eqref{eq:DiscreteAmpStrobo}, one can arrive at the desired result:
      \begin{align}
        \label{eq:FDAZenoLargeN}
        \FDA_n
        \AsymEq &
        -
        \Sum{l=1}{w-1} 
        \frac{
          \frac{1}{\tau} v_\PSI(\mathfrak{s}_l) - \frac{\ip{\PsiDet}{\PsiIn}}{2}
        }{
          \frac{\dd }{\dd z} u_\FDA(z) |_{z=e^{-\mathfrak{s}_l\tau}}
        }
        e^{(n+1)\tau \mathfrak{s}_l}
        \\ \AsymEq & \nonumber
        \Sum{l=1}{w-1} 
        e^{-(n+1)\tau [ \tau \lambda_l + i \omega_l ]}
        \times
        \left\{ \begin{aligned}
          - 2 i
          \lambda_l \theta_l \tau
          \qc & \ket{\PsiIn} \ne \ket{\PsiDet} \\
          - 2 \lambda_l \tau^2
          \qc & \ket{\PsiIn} = \ket{\PsiDet}
        \end{aligned} \right.
        .
      \end{align}
      These expressions for $\FDA_n$ are plugged into the delta-comb definition of $\FDP^\FDA(t) = \sSum{n=1}{\infty} \abs{\FDA_n}^2 \delta(t-n\tau)$.
      The first detection attempt remains untouched, but in the large $n$ part, we replace the comb of delta functions with a smooth function by writing 
      $\sSum{n=2}{\infty} \abs*{\FDA_n}^2 \delta(t - n\tau) \AsymEq \abs*{\FDA_{(t-\tau)/\tau}}^2 / \tau$.
      This is equivalent to performing a local average of $\FDP^\FDA(t)$.
      The additional shift is admissible when $\tau$ is small.
      The result is:
      \begin{align}
        \FDP^\FDA(t)
        \AsymEq & 
        \abs{\ip{\PsiDet}{\PsiIn}}^2 \delta(t-\tau)
        + 
        \left\{ \begin{aligned}
          4 \tau \abs{\Sum{l=1}{w-1} \lambda_l \theta_l e^{-t(\lambda_l\tau + i\omega_l)} }^2
          \\ 
          4 \tau^3 \abs{\Sum{l=1}{w-1} \lambda_l e^{-t(\lambda_l\tau + i\omega_l)} }^2
        \end{aligned} \right.
        ,
      \label{eq:FDPZenoStrobo}
      \end{align}
      where the first line is the result for the transition problem and the second line stands for the return problem.
      Integration of the density shows again the equivalence between both approaches.
      For the transition problem, we find the same result as in Eqs.~\eqref{eq:TDPZenoSchroedi} and \eqref{eq:MomentZenoSchroedi}.
      For the return problem, we find $\TDP^\FDA = \TDP^\PSI = 1$ and the correction factor four $\EA{T^m}^\FDA = 4 \EA{T}^\PSI$ to Eq.~\eqref{eq:MomentZenoSchroedi}.
      
      We thus have found the Zeno limit of the first detection time pdf for the stroboscopic detection protocol.
      Just like the Zeno limit of $\FDP^\PSI(t)$, it consists of a ``fast'' part and a ``slow'' part.
      While the fast part in the stroboscopic detection protocol consists of the very first measurement, 
      it takes the form of a quickly decaying exponential in the non-Hermitian setup.
      The fast dynamics cannot be compared, but the slow dynamics can easily be.
      In fact, they yield the exact same density, except for the return problem, where they differ by the factor of four.
      Our Zeno limit reproduces nicely Eq.~\eqref{eq:TDPMomentEquiv}.
      We also recover the quantization of the mean in the return problem for both stroboscopic and non-Hermitian setups.
      
      The just derived pdf is plotted for the benzene ring and for the random Hamiltonian in Fig.~\ref{fig:FDP}(B,R).
      The figure features a rather large value of $\tau = 0.5 \hbar/\gamma$, which has to be compared to the system's internal time scale $\hbar / (E_\text{max} - E_\text{min}) = 0.25 \hbar/\gamma$.
      Although $\tau$ is rather large, the curves match quite well.
      The total detection probability and the moments for these models are plotted in Figs.~\ref{fig:RandStat} and \ref{fig:BenzStat}.
      In all figures, the Zeno limit data is depicted by green circles and describes the stroboscopic data very well for small $\tau$.

\begin{figure}
        \includegraphics[width=0.99\columnwidth]{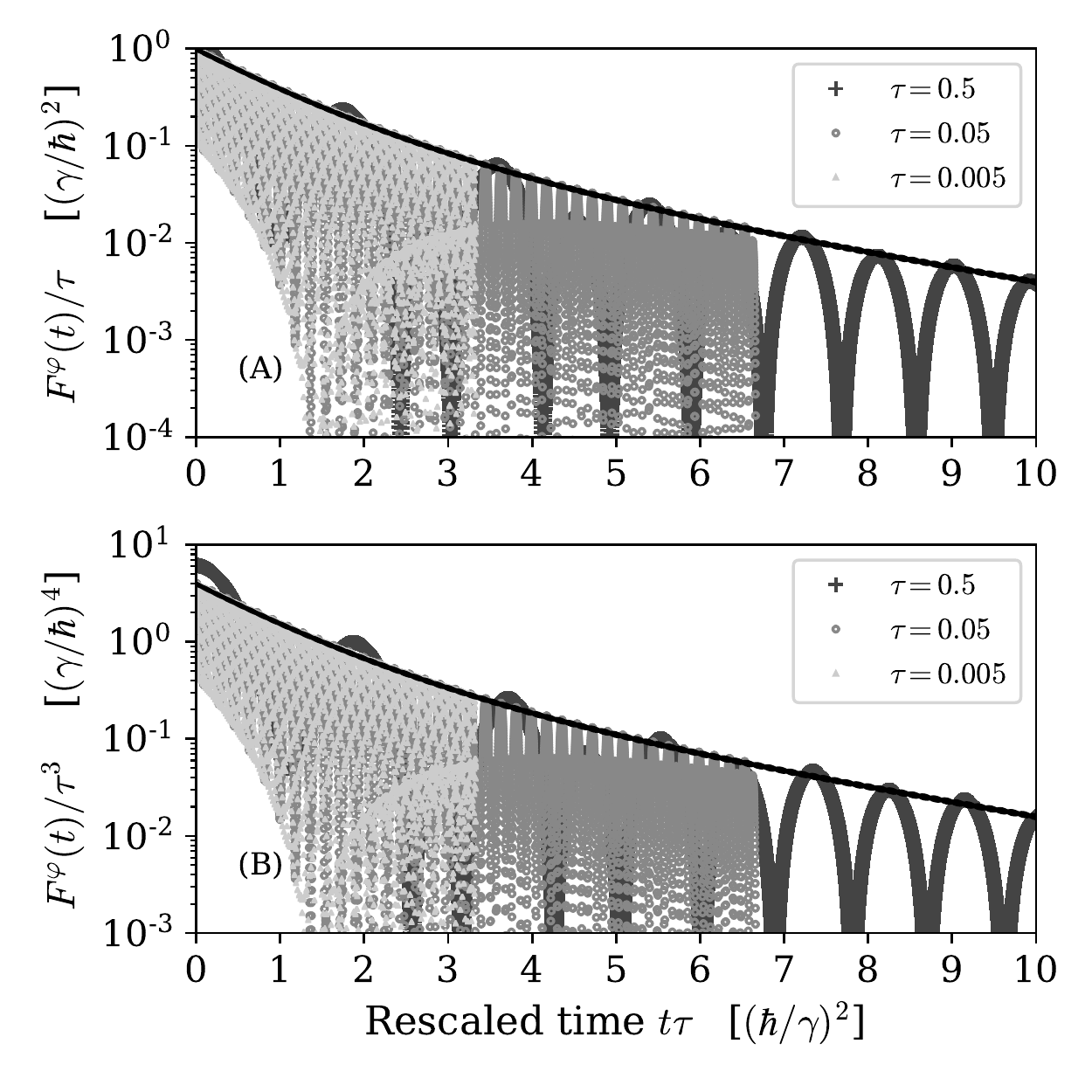}
        \caption{
          Scaling form of the distribution $\FDP^\FDA(t)$ for the benzene ring $\Ham_B$, Eq.~\eqref{eq:BenzHam}.
          Detection and initial states are like in Fig.~\ref{fig:FDP}.
          The distributions' envelopes for different $\tau$ collapse onto each other.
          (Symbols are overlapping; time series are of different lengths.)
          The black solid line is a fit of the envelope.
          There is a different scaling of the prefactor in the return problem (B).
          \label{fig:Collaps}
        }
      \end{figure}

      Furthermore, we find that the slow part of $\FDP^\FDA(t)$ has a very particular scaling in $t\tau$.
      This is best seen in its envelope, when the oscillating terms $e^{i(\omega_l-\omega_{l'})t}$ are neglected.
      This envelope is a scaling function $C(\tau) f(t\tau)$, where the prefactor is either $\tau$ or $\tau^3$,
      depending on whether $\ket{\PsiIn} \ne \ket{\PsiDet}$, or not.
      As a consequence, we can find a data collapse of the pdfs' envelopes of different values for $\tau$.
      This is demonstrated in Fig.~\ref{fig:Collaps}.
      This particular scaling with $t\tau$ was also reported in Refs.~\cite{Muga2008a,Dhar2015a,Thiel2018a}.
      The $t\tau$-scaling in the pdf impacts how the moments vary with $\tau$.
      Namely, we find that $\EA{T^m} \PropTo \tau^{-m}$ for the transition problem and $\EA{T^m} \PropTo \tau^{2-m}$ for the return problem.
      Higher moments diverge as $\tau$ goes to zero, because the dissipation becomes much faster than the internal system dynamics.
      This is a manifestation of the Zeno effect.
      For very small $\tau$, the part of the wave function that was prepared in $\ket{\PsiDet}$, namely $\abs{\ip{\PsiDet}{\PsiIn}}^2$, is immediately detected; 
      this is the meaning of the delta functions in the distributions.
      The remaining amplitude in the system must be transferred to the detection state, but is reflected off it most of the time.
      Detection events that do not take place immediately after preparation are actually very rare and drive the blow-up of the higher moments for small $\tau$.

  \section{In the vicinity of the return problem}
  \label{sec:Return}
    In the preceding sections, we highlighted the special place that the return problem takes amongst all other initial conditions.
    This manifests particularly in the quantized mean first detection time.
    Its behavior switches from linear, $\EA{T}^\FDA \PropTo \tau$, to diverging, $\EA{T}^\FDA \PropTo \tau^{-1}$, 
    depending on the initial state.
    Clearly, the return problem is on a somewhat delicate balance, which is easily perturbed by small alterations of the initial state 
    or by imperfections in the detection protocol.
    In this section, we explore the sensitivity or robustness of the return problem, and in what 
    sense the non-Hermitian and Zeno limit are applicable.

    \subsection{Robustness of the return problem}
      \begin{figure}
        \includegraphics[width=0.9\columnwidth]{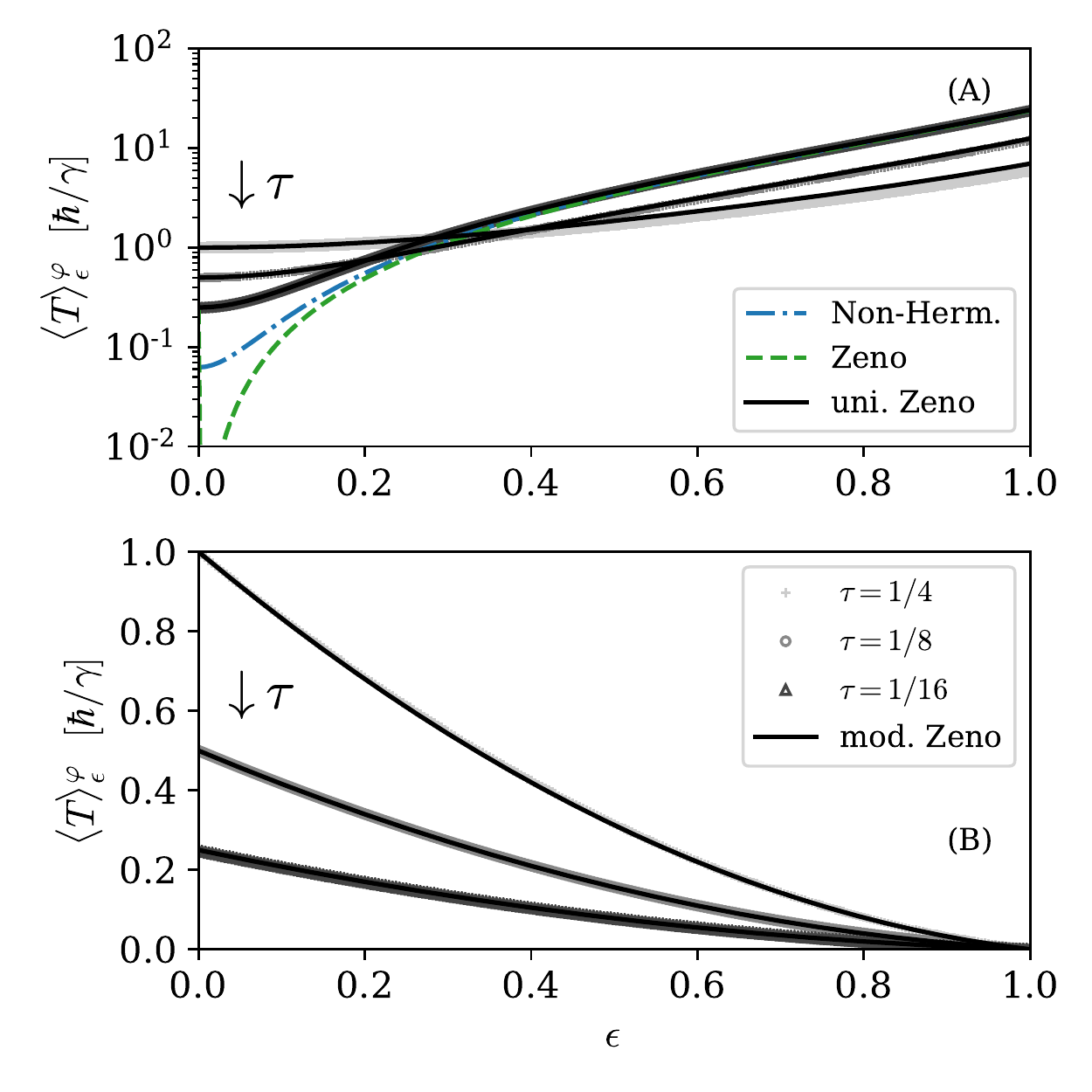}
        \caption{
          Perturbation of the return problem in the benzene ring with $\ket{\PsiDet} = \ket{0}$.
          (A) Initial state is $\ket{\PsiIn^\epsilon} := \epsilon \ket{1} + \sqrt{1 - \epsilon^2} \ket{0}$.
          Gray squares give the stroboscopic data for $\tau = 1/4, 1/8,$ and $1/16 \hbar/\gamma$ (from top to bottom on the left hand side).
          Dashed and dash-dotted lines give the Zeno approximation \eqref{eq:MomentZenoSchroedi} and the non-Hermitian data result \eqref{eq:TDPMomentEquiv}
          for $\tau = 1/16 (\hbar/\gamma)$.
          Both depart from the stroboscopic result close to the return problem (small $\epsilon$).
          The uniform Zeno result (solid black line, Eq.~\eqref{eq:UniformZeno}) matches the stroboscopic data perfectly for small $\tau$.
          (B) Variation of the stroboscopic detection protocol.
          Solid line is the modified Zeno result. 
          \label{fig:Pert}
        }
      \end{figure}
      How resilient are the return statistics to small changes in the initial state?
      When the initial state is equal to, or very close to the detection state, most detection events will 
      occur at $n=1$, shortly after preparation.
      To explore this regime, we consider the mean first detection time $\EA{T}^\FDA_\epsilon$ for an
      initial state of the form
      \begin{equation}
        \ket{\PsiIn^\epsilon}
        :=
        \sqrt{1-\epsilon^2}\ket{\PsiDet} + \epsilon \ket{\PsiIn}
        ,
      \label{eq:Pert1PsiIn}
      \end{equation}
      with, obviously, $0 = \ip*{\PsiIn}{\PsiDet}$.
      When $\epsilon$ vanishes, this initial state describes the return problem.
      As $\epsilon$ increases, we move towards the transition problem.

      For small $\tau$, the mean first detection time is a good observable to describe the contrast between transition and return problems, because of its high sensitivity.
      Fig.~\ref{fig:Pert}(A) shows the $\epsilon$-dependence of the mean for different values of $\tau$
      in the benzene ring.
      It also shows the non-Hermitian result $\EA{T}^\PSI$, which nicely describes the stroboscopic data 
      for large $\epsilon$, but which settles at a fourth of the stroboscopic value when $\epsilon$ 
      goes to zero, nicely demonstrating the necessity of the correction factor in the return problem.
      Furthermore, we plotted the Zeno approximation of Eq.~\eqref{eq:MomentZenoSchroedi} that also matches
      the stroboscopic data for large $\epsilon$, but completely fails to describe $\EA{T}^\FDA_\epsilon$ for small $\epsilon$. 

      When $\epsilon$ and $\tau$ are simultaneously small they compete with each other, and the limits
      $\epsilon\to0$ and $\tau\to0$ do not commute.
      This is why the non-Hermitian and Zeno limits cannot agree for small $\epsilon$, as they are derived by taking $\tau\to0$ first.
      Still, using asymptotic matching, one can compute an uniform Zeno limit, that reproduces the stroboscopic data also for small $\epsilon$, see Fig.~\ref{fig:Pert}(A).
      Note that most of the previous sections' machinery still applies, because we only changed the 
      initial state.
      When repeating the calculations from before with the initial state of Eq.~\eqref{eq:Pert1PsiIn}, 
      we need to take care with the following quantity:
      \begin{equation}
        v_\PSI(\mathfrak{s}_l) - \frac{\tau}{2} \ip{\PsiDet}{\PsiIn^\epsilon}
        \AsymEq
        \left\{ \begin{aligned}
          - i \epsilon \theta_l
          \qc & \tau \ll \epsilon \\
          - \tau [  1 + i \frac{\epsilon}{\tau} \theta_l ]
          \qc & \tau \approx \epsilon
        \end{aligned} \right.
        .
      \label{eq:}
      \end{equation}
      Both equations are derived for small $\tau$, but the second alternative holds 
      when $\epsilon$ is also small and comparable to $\tau$.
      With this result we find
      \begin{equation}
        \EA{T^m}^\FDA_\epsilon
        \AsymEq
        \left\{ \begin{aligned}
          \frac{
            \epsilon^2 \frac{m!}{\tau^m}
            \Sum{l=1}{w-1} 
            \frac{ 2 \lambda_l  \abs{\theta_l}^2 }{ (2\lambda_l)^m }
          }{
            \abs{\ip{\PsiDet}{\PsiIn^\epsilon}}^2
            +
            \epsilon^2
            \Sum{l=1}{w-1} 
            2\lambda_l \abs{\theta_l}^2 
          }
          \qc & \tau \ll \epsilon \\
          \tau^m +
          \frac{m!}{\tau^{m-2}}
          \Sum{l=1}{w-1} 
          \frac{ 2 \lambda_l  \abs{1 + i\frac{\epsilon}{\tau}\theta_l}^2 }{ (2\lambda_l)^m }
          \qc & \tau \approx \epsilon 
        \end{aligned} \right.
        .
      \label{eq:}
      \end{equation}
      So, when $\tau$ is much smaller than the distance $\epsilon$ between initial and detection state, the moments resemble those of the transition problem.
      When $\tau$ and $\epsilon$ are both small, we obtain an interpolation between the return and transition problems.

      A uniform Zeno limit is achieved by the technique of asymptotic matching that combines both lines via:
      $\EA{T}_{\epsilon, \text{ uni}} \AsymEq \EA{T}_{\text{large } \epsilon} + \EA{T}_{\text{small } \epsilon} - \lim_{\epsilon\to0} \EA{T}_{\text{large } \epsilon}$,
      where $\lim_{\epsilon\to0}\EA{T}_{\text{large }\epsilon}$ is the small $\epsilon$ expansion of the first line.
      \begin{align}
        \label{eq:UniformZeno}
        \EA{T^m}_{\epsilon\text{, uni}}^\FDA
        \AsymEq &
        \tau^m +
        \frac{m!}{\tau^{m-2}} \Sum{l=1}{w-1}
        \frac{2\lambda_l}{(2\lambda_l)^m} \bigg\{
          \abs{1+ i\frac{\epsilon}{\tau}\theta_l}^2
          +
        \\ & +
          \frac{\epsilon^2 \abs{\theta_l}^2}{
            \abs{\ip{\PsiDet}{\PsiIn^\epsilon}}^2
            +
            \epsilon^2
            \Sum{l=1}{w-1} 
            2\lambda_l \abs{\theta_l}^2 
          }
          -\epsilon^2 \abs{\theta_l}^2
        \bigg\}
        .
        \nonumber
      \end{align}
      This equation is used in Fig.~\ref{fig:Pert}(A) for $m=1$, which matches the numerical data almost perfectly for small $\tau$.
      The main conclusion we can draw from this example is that the Zeno limit close to the return problem must be carefully 
      performed, because the limits $\ket{\PsiIn}\to\ket{\PsiDet}$ and $\tau\to0$ do not commute.

    \subsection{Robustness of the detection protocol}
      In the same spirit as before, we can ask the question of how stable the return problem is to 
      small disturbances in the detection protocol.
      We now consider a  shift of the first detection epoch by $\epsilon\tau$, with $0 < \epsilon < 1$, 
      such that detection is attempted at $(1-\epsilon)\tau, (2-\epsilon) \tau, \hdots$.
      This scheme is an interpolation between our stroboscopic detection protocol ($\epsilon=0$)
      and the scheme considered in Ref.~\cite{Krovi2006a}, where the first detection attempt
      occurs directly after preparation.

      Again we investigate the mean $\EA{T}^\FDA_\epsilon$ in the return problem.
      In the return problem, we find $\EA{T}^\FDA_{\epsilon=0} = w\tau$, but $\EA{T}^\FDA_{\epsilon=1} = 0$,
      because the system is detected directly after preparation.
      As $\epsilon$ varies, one interpolates between both results continuously, as depicted in 
      Fig.~\ref{fig:Pert}(B) for different values of $\tau$.

      A formula for $\EA{T^m}^\FDA_\epsilon$ in the Zeno limit is easily found.
      Reworking the argument in Sec.~\ref{sec:MagicFactor}, we have that the probability of surviving the first measurement is now reduced by a factor $(1-\epsilon)^2$, since less probability has been transferred off the detection site before measurement.
      Similarly, the modified amplitude of first successful detection at the $n$-th attempt ($n>1$) is reduced by the same factor, as are all the moments of the first detection time.
      The mean  is plotted in Fig.~\ref{fig:Pert}(B) where it agrees well with the numerical data.
      The result is an interpolation between the usual return result, namely four times Eq.~\eqref{eq:MomentZenoSchroedi}, and zero.
      For the first moment this gives: $\EA{T}^\FDA_\epsilon \AsymEq (1 - \epsilon)[\epsilon + w (1-\epsilon) ] \tau$, as $\tau\to0$.
      For the perturbed detection protocol, there is no competition between $\epsilon$ and $\tau$.

  \section{Summary and Discussion}
  \label{sec:SumDisc}
    The quantum first detection problem assesses the statistics of the first successful of many 
    repeated detection attempts in the state $\ket{\PsiDet}$ performed stroboscopically with frequency $1/\tau$.
    The non-Hermitian Schr\"odinger equation \eqref{eq:Schroedi} is an alternative model, where the projective measurements are replaced by an imaginary potential on $\ket{\PsiDet}$.
    We presented the formal solutions to both problems and compared them with each other in the limit of small $\tau$.
    It was demonstrated that they then both yield the same statistics, except in the return problem, $\ket{\PsiIn} = \ket{\PsiDet}$,
    which necessitates a correction to restore the equivalence.
    This was also shown in extensive numerical simulations.
    For systems with a discrete energy spectrum, we presented another formal solution in terms of the poles $\mathfrak{z}_l$ of 
    the generating function $\FDA(z)$ or the poles $\mathfrak{s}_l$ of the Laplace transformed wave function $\PSI(s)$.
    The poles can be obtained from the stationary points of an electrostatic potential in both situations.
    It was demonstrated that the mean first detection time $\EA{T}$ is quantized for the return problem, re-deriving the result of \cite{Gruenbaum2013a}.
    The relevant integer is $w$, the number of energy levels that appear in the spectral decomposition of $\ket{\PsiDet}$.
    Using the electrostatic analogy, we found the Zeno limit $\tau\to0$ of the non-Hermitian description.
    Finally, using the Zeno limit, we analyzed the stroboscopic detection protocol in the vicinity of the return problem.

    Throughout this article, we considered the simultaneous limit $\tau\to0$ and $n\to\infty$.
    Our technique here was anticipated in Ref.~\cite{Thiel2018a}.
    This way, we avoided the trivial result $\FDA_n \AsymEq \delta_{n,1} \abs{\ip{\PsiDet}{\PsiIn}}^2$ from Eq.~\eqref{eq:DefFDA},
    where all dynamical information is lost.
    Still, as is evident from the nature of the limit, we can not map the region $t \ApproxEq 0$, 
    where $n = t/\tau$ is actually not large.
    The discrepancy between the fast parts in Eqs.~\eqref{eq:FDPZenoSchroedi} and \eqref{eq:FDPZenoStrobo} are symptoms of this inability.
    This discrepancy in the fast dynamics can also be expected from how the two models behave for small times as addressed in Sec.~\ref{sec:MagicFactor}.
    Still, when comparing the actual probabilities of early absorption from Eqs.~\eqref{eq:FDPZenoSchroedi} and \eqref{eq:FDPZenoStrobo}, we find pretty good agreement: 
    The relative error between $\sInt{0}{\tau}{t} \FDP^\PSI(t)$ and $\sInt{0}{\tau}{t} \FDP^\FDA(t)$ is approximately $2\%$.
    
    Our approximation scheme $z = e^{-s\tau}$ with consecutive small $\tau$ expansion is similar 
    to the Tustin or bi-linear transformation in signal theory \cite{Fadali2013a}.
    This becomes clearer, when $z = e^{-s\tau} \AsymEq (1 - \frac{s\tau}{2})/(1 + \frac{s\tau}{2})$ is replaced 
    by its Pad\'e approximation.
    The Tustin transform is used to transform continuous-time filters into discrete-time ones and vice versa.
    It captures the small frequency behavior correctly, but distorts the high-frequencies; 
    a phenomenon known as ``frequency warping'', that makes resonant detection periods impossible 
    in the non-Hermitian limit.
    The first resonant detection period, defined by $\tau_c = 2\pi \hbar/(E_\text{max} - E_\text{min})$, therefore poses a hard limit for the validity of the non-Hermitian description.
    This is well supported by our numerical data.

    Furthermore, when we derived $\TDP^\FDA$ in the Zeno limit in Eq.~\eqref{eq:TDPZenoSchroedi}, we encountered $\abs{\ip{\PsiDet}{\PsiIn}}^2$.
    This is the probability of detection directly after preparation.
    $\TDP^\FDA(\PsiIn) - \abs*{\ip{\PsiDet}{\PsiIn}}^2 = \sSum{l=1}{w-1} 2\lambda_l \abs{\theta_l}^2$ 
    is the difference in the total detection probability between the stroboscopic detection protocol 
    and a ``one-shot'' detection protocol.
    In Ref.~\cite{Thiel2019c,Thiel2019d}, we demonstrated that this quantity can be bounded by an uncertainty relation,
    so that:
    \begin{equation}
      \TDP^\FDA(\PsiIn) - \abs{\ip{\PsiDet}{\PsiIn}}^2 
      = 
      \Sum{l=1}{w-1} 2\lambda_l \abs{\theta_l}^2
      \ge
      \frac{\abs*{\mel*{\PsiDet}{\comm*{\Ham}{\Detect}}{\PsiIn}}^2}{\sVar{\Ham}_{\PsiDet}}
      ,
    \label{eq:}
    \end{equation}
    where $\Detect := \dyad{\PsiDet}$ and $\sVar{\Ham}_{\PsiDet} := \ev*{\Ham^2}{\PsiDet} - [\ev*{\Ham}{\PsiDet}]^2$ are the energy fluctuations in the 
    detection state.

    The special character of the return problem -- in particular the quantization of the mean first return time --
    was already discussed and recognized for the stroboscopic detection protocol \cite{Gruenbaum2013a, Sinkovicz2016a, Friedman2017a}.
    For the non-Hermitian Schr\"odinger equation, the quantization does not seem to have been previously noted.
    It will be showcased in a separate publication \cite{Thiel2019f}.

    We have demonstrated that the analogy between the non-Hermitian Sch\"odinger equation and the stroboscopic
    detection protocol is very delicate.
    The equivalence of both depends crucially on the value of $\tau$,
    the exact definition of the detection protocol, and the initial state in question.
    In the vicinity of the return problem, both the Zeno and non-Hermitian approximation are particularly untrustworthy.
    This shows that the first detection statistics may be quite sensitive to
    their exact operational definition.
    The popular non-Hermitian description must be motivated with great care to detail in any repeated measurement setup.

  \begin{acknowledgments}
    Felix Thiel thanks DFG (Germany) to support him under grant no. TH 2192/1-1 and TH 2192/2-1.
    The support of Israel Science Foundation's grant 1898/17 is acknowledged.
  \end{acknowledgments}

  \appendix
  \section{Adiabatic elimination of the fast mode}
  \label{app:Adiabatic}
    Following Dhar et al. \cite{Dhar2015b}, we show in this section how to obtain another non-Hermitian equation with a small optical potential from Eq.~\eqref{eq:Schroedi}.
    This is achieved via adiabatic elimination of the fast mode.
    Starting from Eq.~\eqref{eq:Schroedi}, we decompose the wave function into two orthogonal parts $\ket{\psi(t)} = \ket{\PsiDet} \PSI(t) + \ket*{\bar{\psi}(t)}$, such that $\Detect\ket{\bar{\psi}(t)} = 0$.
    We assume that the initial state has no overlap with $\ket{\PsiDet}$, $\ip{\PsiDet}{\PsiIn} = 0$.
    After Laplace transformation, the Schr\"odinger equation reads in block form:
    \begin{equation}
      \left\{ \begin{aligned}
        i s \PSI(s) 
        =
        [ \ev*{\Ham} - \frac{2i}{\tau} ] \PSI(s)
        + \bra{\PsiDet}\Ham (\Id - \Detect) \ket{\bar{\psi}(s)}
        \\
        i [ s \ket{\bar{\psi}(s)} - \ket{\PsiIn} ]
        =
        (\Id - \Detect) \Ham \ket{\PsiDet} \PSI(s)
        + \Ham_\text{Z} \ket{\bar{\psi}(s)}
        ,
      \end{aligned} \right.
    \label{eq:}
    \end{equation}
    where $\ev*{\Ham} = \ev*{\Ham}{\PsiDet}$ and $\Ham_\text{Z} := (\Id - \Detect) \Ham (\Id - \Detect)$ is the Zeno Hamiltonian, see \cite{Facchi2008a}.
    Solving the first equation for $\PSI(s)$ and plugging the result into the second equation yields:
    \begin{equation}
      i [ s \ket{\bar{\psi}(s)} - \ket{\PsiIn} ]
      =
      \qty[
        \Ham_\text{Z} 
        + \frac{ \Ham_1 }{is - \ev*{\Ham} + \frac{2i}{\tau} }
      ]\ket{\bar{\psi}(s)}
      ,
    \label{eq:}
    \end{equation}
    where $\Ham_1 = (\Id - \Detect)\Ham \Detect\Ham(\Id-\Detect)$.
    When $\tau$ is very small, the terms $is - \ev*{\Ham}$ can be neglected in the denominator and we obtain an effective non-Hermitian Hamiltonian that only acts on the subspace $(\Id - \Detect)$.
    An inverse Laplace transform gives the effective Schr\"odinger equation:
    \begin{equation}
      i \hbar \dv{t} \ket{\bar{\psi}(t)}
      =
      \qty[ 
        \Ham_\text{Z} 
        - i \frac{\tau}{2\hbar} 
        \Ham_1
      ] \ket{\bar{\psi}(t)}
      .
    \label{eq:}
    \end{equation}
    This is exactly the equation used in Refs.~\cite{Dhar2015a, Dhar2015b, Lahiri2019a, Elliott2016a}.

  \section{Lazy detector limit}
  \label{app:Lazy}
    In sec.~\ref{sec:Ultimate} we derived the Zeno limit for the non-Hermitian Schr\"odinger equation.
    This was achieved by a perturbation of the equation $0 = 1 + 2u_\PSI(s)/\tau$ as $\tau\to0$.
    A similar procedure is possible in the opposite limit $\tau\to\infty$, when the detector becomes slower and slower.
    Obviously, this has no correspondence with the stroboscopic detection protocol, whence we omitted its discussion in 
    the main text.
    Nevertheless, it is clearly justified as a proper non-Hermitian system with a very weak dissipation term,
    which is why we present it here.
    The general considerations of sec.~\ref{sec:Discrete} still hold.
    So, it is only necessary to find the poles $\bar{\mathfrak{s}}_l$ in this limit.

    When $\tau$ becomes large in $0=1+2u_\PSI(\mathfrak{s})/\tau$, $u_\PSI(\mathfrak{s})$ must become large as well to satisfy the equation.
    For this reason, we expand $u_\PSI(s)$ around its singularities $s = -i E_l / \hbar$, by making the ansatz
    $\bar{\mathfrak{s}}_l \AsymEq - i (E_l / \hbar) - (a_l/\tau)$, for $l=1, \hdots, w$.
    Plugging this ansatz into Eq.~\eqref{eq:DefPolesSchroedi} together with Eq.~\eqref{eq:DiscResSchroedi} reveals $a_l = p_l$.
    That means we find 
    \begin{equation}
      \bar{\mathfrak{s}}_l
      \AsymEq
      - \frac{2p_l}{\tau} - i\frac{E_l}{\hbar}
      \qc l = 1, \hdots, w
      .
    \label{eq:LazyPoles}
    \end{equation}
    The same result is obtained when one applies regular perturbation theory to find the eigenvalues $\tilde{E}_l(\tau) = - i \hbar \mathfrak{s}_l$ of Eq.~\eqref{eq:Schroedi}.

    Plugging these poles into the functions $v_\PSI(s)$ and $u_\PSI(s)$, we obtain in leading order:
    \begin{equation}
      v_\PSI(\bar{\mathfrak{s}}_l)
      \AsymEq
      q_l
      \qc 
      u_\PSI'(\bar{\mathfrak{s}}_l)
      \AsymEq
      - \frac{\tau^2}{4p_l}
      .
    \label{eq:}
    \end{equation}
    With these we obtain $\PSI(s)$ from Eq.~\eqref{eq:DiscreteAmpSchroedi}:
    \begin{equation}
      \PSI(t)
      \AsymEq
      \Sum{l=1}{w} p_l q_l 
      e^{-t [ \frac{2p_l}{\tau} + i \frac{E_l}{\hbar} ] }
      .
    \label{eq:LazyAmp}
    \end{equation}
    Integration of $\FDP^\PSI(t) = 4 \abs{\PSI(t)}^2/\tau$ yields the total detection probability and the moments.
    In leading order in $\tau\to\infty$, these are independent of the energy levels $E_l$
    and just depend on the charges $p_l$ and $q_l$.
    \begin{align}
      \TDP
      \AsymEq &
      \Sum{l=1}{w} p_l \abs{q_l}^2
      \\
      \EA{T^m}
      \AsymEq &
      m! \frac{\tau^m}{4^m}
      \frac{
        \sSum{l=1}{w} p_l \abs{q_l}^2 p_l^{-m}
      }{
        \sSum{l=1}{w} p_l \abs{q_l}^2
      }
      .
    \end{align}
    It is not obvious that the here derived normalization for $\tau\to\infty$ coincides with the $\tau\to0$ limit of Eq.~\eqref{eq:TDPZenoSchroedi} from the main text.
    However, judging from our numerical simulations depicted in Figs.~\ref{fig:RandStat} and \ref{fig:BenzStat}, which show perfect constancy in $\TDP^\PSI(\tau)$,
    we can conclude that they are the same.
    Furthermore, it correctly reproduces the exact result for the stroboscopic detection protocol of Ref.~\cite{Thiel2019a}.
    We also reproduce the quantization of the return problem for large $\tau$: $\EA{T} = w \tau/4$.

    Another remark on Eq.~\eqref{eq:LazyAmp} is called for.
    In contrast to the small $\tau$-expansion of the main text, we find the unified scaling $\FDP^\PSI(t) = \tau^{-1} f(t/\tau)$ for the envelope here.
    There is no separation between a fast and some slow modes.
    All modes' time scales are of the same order of magnitude $\tau$.

  \section{Calculations for the infinite line}
  \label{app:InfLine}
    In this section, we explain how all quantities pertaining to the infinite line Hamiltonian \eqref{eq:InfLineHam} have been obtained.
    To simplify our equations, we work in units of time where $\hbar/\gamma = 1$.
    Furthermore, we will use the abbreviation $\Gamma = 2/\tau$ when convenient.
    We will make heavy use of the techniques of Krapivsky, Luck and Mallick described in Ref.~\cite{Krapivsky2014a}, repeating some of their calculation, but also expanding upon them.

    Ref.~\cite{Friedman2017b} reported the transition amplitudes for this model:
    \begin{equation}
      \mel*{x}{\TEO(t)}{y}
      =
      i^{\abs{x-y}} \BesselJ{\abs{x-y}}{2 t}
      .
    \label{eq:app:InfLineTransAmp}
    \end{equation}
    Here $\BesselJ{n}{x}$ is the Bessel function of the first kind and $\TEO(t) = e^{-it\Ham/\hbar}$.
    Henceforth, we write $\xi = \abs{x-y}$.
    This expression is plugged into $ [s + i \Ham/\hbar]^{-1} = \sInt{0}{\infty}{t} e^{- st - i t\Ham/\hbar}$ to obtain the resolvent of the Hamiltonian.
    \begin{align}
      \mel*{x}{\frac{1}{s + \frac{i}{\hbar}\Ham}}{y}
      = &
      \frac{
        \qty[
          \tfrac{i}{2}\qty(
            \sqrt{4 + s^2} - s
          )]^\xi
      }{
        \sqrt{4 + s^2}
      }
      ,
    \label{eq:}
    \end{align}
    where Eq.~6.611.1 of \cite[p. 694]{Gradshteyn2007a} was used.
    We consider the detection state $\ket{\PsiDet} = \ket{0}$ and the initial state $\ket{\PsiIn} = \ket{\xi}$.
    Therefore, above expression gives $v_\PSI(s)$ and also $u_\PSI(s)$ upon setting $\xi =0$.
    This results in:
    \begin{equation}
      \PSI_\xi(s)
      =
      \frac{
        \qty[
          \tfrac{i}{2}\qty(
            \sqrt{4 + s^2} - s
        )]^\xi
      }{
        \Gamma
        + \sqrt{4 + s^2}
      }
      .
    \label{eq:AppPSIInfLine}
    \end{equation}
    The initial state is carried in a subscript from here on.

    We will first find an expression for $\PSI(t)$ in time domain and then proceed to compute the first moment of $\EA{T}^\PSI$.

    \subsection{Wave function in time domain}
      In the return problem, $\xi=0$, $\PSI_0(s)$ is a function of $\sqrt{1+ s^2/4}$ only.
      By virtue of Eq.~1.1.1.37 of \cite[p. 6]{Prudnikov1992a}, we thus find:
      \begin{equation}
        \PSI_0(t)
        =
        e^{-\Gamma t}
        -
        2 t 
        \Int{0}{1}{y}
        \BesselJ{1}{2ty}
        e^{- \Gamma\sqrt{1-y^2}}
        .
      \label{eq:app:PSI0}
      \end{equation}
      The remaining integral can be obtained numerically.
      
      For the transition case, $\xi\ne 0$, we use Eq.~2.9.1.15 of \cite[p. 47]{Prudnikov1992a} to identify the numerator of Eq.~\eqref{eq:AppPSIInfLine} with $i^{\xi} \xi \BesselJ{\xi}{2t} / (2t)$.
      The denominator is the same expression as before; their product becomes a convolution in 
      time domain:
      \begin{align}
        \PSI_\xi(t)
        =
        \frac{\xi}{2}
        i^{\xi}
        \Int{0}{t}{t'}
        \frac{\BesselJ{\xi}{2(t-t')}}{t-t'}
        \PSI_0(t')
        .
      \label{eq:app:PSIXI}
      \end{align}
      $\PSI_\xi(t)$ can be obtained by numerical quadrature of the last two integrals.
      
      As a note, we can take the limit $\tau\to0$ in Eq.~\eqref{eq:app:PSI0} to obtain the much simpler result:
      \begin{equation}
        \PSI_\xi(s)
        \AsymEq
        i^{\xi} 
        \frac{\tau}{2}
          \qty[
            \sqrt{1 + \tfrac{s^2}{4}}
            - \frac{s}{2}
          ]^{\xi}
      \label{eq:}
      \end{equation}
      which transforms to
      \begin{equation}
        \PSI_\xi(t)
        \AsymEq
        i^{\xi}
        \frac{\xi \tau}{2 t}
        \BesselJ{\xi}{2t}
      \label{eq:}
      \end{equation}
      and recovers our result from Ref.~\cite{Thiel2018a}.

    \subsection{Total detection probability}
      From Eq.~\eqref{eq:DefTDPSchroedi} of the main text, the total detection probability and the moments are given by a contour integral.
      However, careful attention must be paid to the branch cuts of the square-root function.
      Eq.~\eqref{eq:DefTDPSchroedi} is derived from the following identity: $\TDP^\PSI = \lim_{\epsilon\searrow0} \sInt{0}{\infty}{t} e^{-\epsilon t} 2\Gamma [\PSI(t)]^* \PSI(t)$.
      Writing $\PSI(t) = \sInt{\Bromwich}{}{s} e^{s t} \PSI(s) / (2\pi i)$ and switching the order of integration, one finds:
      \begin{align}
        \TDP^\PSI \EA{T^m}^\PSI
        =
        \lim_{\epsilon\searrow0} 2 \Gamma 
        \int\limits_\Bromwich\frac{\dd s}{2\pi i}
        \PSI^*(\epsilon - s) \PSI(s)
        ,
      \end{align}
      where $0 < \Re[s] < \epsilon$ so that all intermediary integrals converge.
      We thus parametrize the Bromwich path as $s = \lambda + i \omega$, with $0 < \lambda < \epsilon$, use the definition $\PSI^*(s) = [\PSI(s^*)]^*$ and take the limit $\epsilon\to0$.
      This shows that both factors in the integrand must be evaluated at $0^+ + i\omega$ and fixes the correct branch of the square-root function.
      \begin{align}
        \TDP^\PSI
        =
        \frac{2 \Gamma }{2 \pi}
        \Int{-\infty}{\infty}{\omega}
        \abs{ \PSI(0^+ + i \omega) }^2
        .
      \end{align}
      The square-root functions in $\PSI(s) = \PSI_\xi(s)$ are replaced by
      \begin{equation}
        \sqrt{4 + (0^+ + i\omega)^2}
        =
        i \Sign{\omega} 
        \sqrt{\omega^2 - 4}
        ,
      \label{eq:SqrtBranch}
      \end{equation}
      for $\abs{\omega} > 2$ and the obvious limit $\sqrt{4 - \omega^2}$ for $\abs{\omega}<2$.
      We abbreviate $\delta := \Sign{\omega}\sqrt{\omega^2-4}$ and $\bar{\delta} := \sqrt{4 - \omega^2}$.
      With this formula, we find that:
      \begin{equation}
        \abs*{\PSI_\xi(0^+ + i\omega)}^2
        =
        \left\{ \begin{aligned}
          \frac{1}{(\Gamma + \bar{\delta})^2}
          \qc & \abs{\omega} \le 2 \\
            \frac{
              \qty[\frac{1}{2} \qty(\omega - \delta) ]^{2\xi}
            }{
              \Gamma^2 + \delta^2
            }
          \qc & \abs{\omega} > 2
        \end{aligned} \right.
      \label{eq:TDPIntegrand}
      \end{equation}
      This expression is integrated over $\omega$ from $-\infty$ to $\infty$ and multiplied by $2\Gamma/(2\pi)$ to yield $\TDP^\PSI$.
      This has already been done in Ref.~\cite{Krapivsky2014a}, whose solution we here cite:
      \begin{widetext}
        \begin{equation}
          \TDP(\tau)
          =
          \left\{ \begin{aligned}
            \frac{2}{\pi} \frac{1}{\tau^2 ( 1 - \tau^2 )} \qty[
              (\pi - 2\tau)(1 - \tau^2)
              + \tau^3
              - [2(1-\tau^2)^2 + \tau^2]
              \frac{\arccos\tau}{\sqrt{1-\tau^2}}
            ]
            \qc & \ket{\PsiIn} = \ket{1} \\
            \frac{1}{\pi}\frac{2\tau}{1 - \tau^2} \qty[
              1 + \frac{1-2\tau^2}{\tau} \frac{\arccos\tau}{\sqrt{1-\tau^2}}
            ]
            \qc & \ket{\PsiIn} = \ket{0} \\
          \end{aligned} \right.
        \label{eq:app:TDPInfLine}
        \end{equation}
      \end{widetext}
      These are the curves plotted in Fig.~\ref{fig:InfLineStat} for the non-Hermitian Schr\"odinger equation.
      The corrected non-Hermitian data was obtained form Eq.~\eqref{eq:app:TDPInfLine} as well via $4\TDP^\PSI - 3$, see Eq.~\eqref{eq:TDPMomentEquiv}.

    \subsection{Mean detection time}
      The mean detection time is computed in a similar manner to before.
      First, we note that the derivative of $\PSI_\xi(s)$ can be written as:
      \begin{equation}
        -\dv{\PSI_\xi(s)}{s}
        =
        \PSI_\xi(s)
        \frac{
          s 
          + \xi \qty(
            \Gamma 
            + \sqrt{4 + s^2}
          )
        }{
          \qty[ 4 + s^2 ]
          +
          \Gamma \sqrt{4 + s^2}
        }
        .
      \label{eq:AppPsiDeriv}
      \end{equation}
      The procedure from before applied to Eq.~\eqref{eq:DefMomentSchroedi} then yields:
      \begin{equation}
        \EA{T}^\PSI 
        = 
        -
        \frac{1}{\TDP^\PSI} \frac{2\Gamma}{2\pi i}
        \Int{-\infty}{\infty}{\omega}
        \left.
        \abs{\PSI_\xi(s)}^2 \dv{\ln \PSI_\xi(s)}{s}\right|_{s=0^++i\omega}
        .
      \label{eq:InfLineMeanIntegral1}
      \end{equation}
      Combining Eqs.~(\ref{eq:SqrtBranch}, \ref{eq:TDPIntegrand}, \ref{eq:AppPsiDeriv}, \ref{eq:InfLineMeanIntegral1}) and using the symmetry of the integrands gives $\EA{T}^\PSI$ as a sum over two integrals: one over $\abs{\omega}<2$ and one over $\abs{\omega}>2$:
      \begin{align}
        \EA{T}^\PSI
        = & \nonumber
        \frac{2\Gamma}{2\pi\TDP^\PSI} \bigg\{
          \Int{0}{2}{\omega} \frac{2\xi}{\bar{\delta} (\Gamma + \bar{\delta})^2}
        + \\ & +
          \Int{2}{\infty}{\omega} \frac{2 \Gamma \omega \qty[\frac{1}{2} ( \omega - \delta )]^{2\xi}}{\delta (\Gamma^2 + \delta^2)^2}
        \bigg\}
        .
      \label{eq:}
      \end{align}
      We call the integral in the first line $I_1$ and the one in the second line $I_2$.

      The first integral is solved by changing variables to $\delta$ with $\dd \delta = \omega \dd \omega / \delta$ and using Mathematica:
      \begin{equation}
        I_1 
        =
        \frac{1}{2\Gamma}
        \frac{2\xi \tau}{1 - \tau^2}
        \qty[
          \frac{\arccos(\tau)}{\sqrt{1-\tau^2}}
          - \tau
        ]
        .
      \label{eq:FirstInt}
      \end{equation}
      For the second integral, we use the variable transform $\omega = 2 \cosh x$, such that $\delta = 2 \sinh x$, $\omega-\delta = 2 e^{-x}$, and $\dd \omega = 2 \sinh x \dd x$.
      This gives:
      \begin{equation}
        I_2 
        = 
        \Int{2}{\infty}{\omega} \frac{2 \Gamma \omega \qty[\frac{1}{2} ( \omega - \delta )]^{2\xi}}{\delta (\Gamma^2 + \delta^2)^2}
        =
        \frac{\tau^2}{2\Gamma}
        \Int{0}{\infty}{x} \frac{e^{-2 \xi x}\cosh x }{(1 + \tau^2\sinh^2 x)^2}
        ,
      \label{eq:}
      \end{equation}
      where we already replaced $2/\Gamma = \tau$ for convenience.
      The exact integral is calculated with Mathematica and results in a complicated mix of polynomial and logarithmic terms in $\tau$ for general $\xi$.
      For $\xi=0$ and $\xi=1$ the result is:
      \begin{equation}
        I_2 
        =
        \left\{ \begin{aligned}
          \frac{1}{2\Gamma} \frac{\pi}{4} \tau \qc & \xi = 0 \\
          \frac{1}{2\Gamma} \frac{\pi}{4} \frac{2 + \tau^2}{\tau} - 1 - \frac{\arccos(\tau)}{\tau \sqrt{1 - \tau^2}}
          \qc & \xi = 1 \\
        \end{aligned} \right.
        .
      \label{eq:}
      \end{equation}
      Adding $I_1$ and multiplying by $2\Gamma / (2\pi \TDP^\PSI)$ gives the conditional mean detection time:
      \begin{widetext}
        \begin{equation}
          \EA{T}^\PSI 
          =
          \left\{ \begin{aligned}
           \frac{1}{\TDP^\PSI(\xi=1)}\qty{
              \frac{2 - 3 \tau^2 + \tau^2}{8\tau (1-\tau^2)^2}
              - \frac{1 + \tau^2}{2\pi(1 - \tau^2)}
              + \frac{1-3\tau^2}{2\pi \tau} \frac{\arccos(\tau)}{\sqrt{1-\tau^2}}
            }
            \qc & \ket{\PsiIn} = \ket{1} \\
            \frac{1}{\TDP^\PSI(\xi=0)} \frac{\tau}{8}
            \qc & \ket{\PsiIn} = \ket{0}
          \end{aligned} \right.
        \label{eq:app:MeanTInfLine}
        \end{equation}
      \end{widetext}
      Combined with Eq.~\eqref{eq:app:TDPInfLine} this result is plotted in Fig.~\ref{fig:InfLineStat} of the main text.

    \subsection{A note on the total detection probability}
      The total detection probability for the NHH is given by Eq.~\eqref{eq:app:TDPInfLine}.
      Expanding the result for $\xi=0$ for small $\tau$ one obtains:
      \begin{align}
        \TDP^\PSI
        &= 
        1 
        - \frac{1}{4} \tau^2
        + \frac{8}{3\pi}\tau^3 
        - \frac{9}{8}\tau^4 
        + \frac{64}{15\pi}\tau^5
        - \frac{25}{16}\tau^6 
        + \ldots
      \end{align}
      The first detection amplitudes from the stroboscopic approach are obtained from the renewal equation and Eq.~\eqref{eq:app:InfLineTransAmp}.
      The first terms read:
      \begin{equation}
        \FDA_n
        = 
        \left\{ \begin{aligned}
          \BesselJ{0}{2\tau} \qc & n=1 \\ 
          - \frac{2\tau}{n} J_1(2n\tau)
          - \frac{3\tau^2}{n^2} J_2(2n\tau)
          + \ldots 
          \qc & n>1 
        \end{aligned} \right.
      \end{equation}
      From this and $\TDP^\FDA = \sSum{n=1}{\infty} \abs{\FDA_n}^2$, we can compute the total detection probability in orders of $\tau$:
      \begin{equation}
        \TDP^\FDA
        = 
        1 
        - 2\tau^2 
        + \frac{32\tau^3}{3\pi} 
        - \frac{9\tau^4}{2} 
        + \frac{256\tau^5}{15\pi} 
        - \frac{50\tau^6}{9} 
        + \ldots
      \end{equation}
      Comparing the expressions for the stroboscopic and the NHH approach, we see that, surprisingly, the ``corrected" NHH result $4\TDP^\PSI-3$ agrees with the stochastic result for the first five orders in $\tau$, disagreeing only at order $\tau^6$.

  \section{Details of the simulations}
  \label{app:Numerics}
    In this section, we describe how the data for the figures was obtained.
    \subsection{Infinite line}
      \paragraph{Total detection probability and mean first detection time}
        The total detection probability that is plotted in Fig.~\ref{fig:InfLineStat} was generated in the following way.
        $\TDP^\PSI(\tau)$ was already computed in Ref.~\cite{Krapivsky2014a} for the infinite line.
        We repeated the result in Eq.~\eqref{eq:app:TDPInfLine}, and plotted these curves in Fig.~\ref{fig:InfLineStat}.
        The corrected non-Hermitian data was obtained form Eq.~\eqref{eq:app:TDPInfLine} as well via $4\TDP^\PSI - 3$, see Eq.~\eqref{eq:TDPMomentEquiv}.
        The same was done for the mean, that we calculated in Eq.~\eqref{eq:app:MeanTInfLine}.

        The total detection probability and the mean first detection time for the stroboscopic detection protocol were obtained from the renewal equation \eqref{eq:QuantumRenewal} and the exact expression \eqref{eq:app:InfLineTransAmp}.
        $\TDP^\FDA(\tau)$ is approximated by the sum $\sSum{n=1}{N} \abs{\FDA_n}^2$, where $N$ is chosen such that the last summand is sufficiently small.
        The same approach was taken to compute $\EA{T} \ApproxEq \sSum{n=1}{N} \abs{\FDA_n}^2 (n\tau) / \TDP^\FDA(\tau)$.
        This way the curves for the stroboscopic data in Fig.~\ref{fig:InfLineStat} were generated.

      \paragraph{Probability density function}
        The stroboscopic data in Fig.~\ref{fig:FDP}(L) was generated from the renewal equation as explained above.
        Eq.~\eqref{eq:app:InfLineTransAmp} was used together with Eq.~\eqref{eq:QuantumRenewal} to obtain $\FDA_n$.
        Still $\FDP^\FDA(t)$ contains the $\delta$-functions $\delta(t - n\tau)$.
        To avoid them, we plotted the ``local average'' $(1/\tau) \sInt{(n-1/2)\tau}{(n+1/2)\tau}{t} \FDP^\FDA(t) = \abs{\FDA_n}^2/\tau$ instead of $\FDP^\FDA(n\tau)$.
        Hence, we used the data points $(n\tau, \abs{\FDA_n}^2/\tau)$ for the stroboscopic data, where $\FDA_n$ was obtained as described above.

        The non-Hermitian data was obtained from numerical quadrature of Eqs.~\eqref{eq:app:PSI0} and \eqref{eq:app:PSIXI} that we derived above and from $F(t) = (4/\tau)\abs{\PSI(t)}^2$ .

    \subsection{Benzene Ring}
      The ring Hamiltonian $\Ham_B$ has four distinct energy levels that have overlap with $\ket{\PsiDet}$ and therefore $w=4$.
      Both Hamiltonians have four energy levels which have overlap with $\ket{\PsiDet}$ and therefore $w=4$.
      The resolvents $u_\PSI(s)$ and $u_\FDA(s)$ are found symbolically from the matrix representations \eqref{eq:BenzHam} and from Eq.~\eqref{eq:DefPsiDetEx} using Mathematica.
      Since $w$ is small enough, the poles $\mathfrak{z}_l$ and $\mathfrak{s}_l$ can also be determined symbolically, as none of the polynomials encountered have order larger than four.

      \paragraph{Probability density function}
        Using the exact expressions of the poles, the resolvents and of $v_\PSI(s)$ as well as $v_\FDA(z)$, we can find $\FDA_n$ and $\PSI(t)$ from Eqs.~\eqref{eq:DiscreteAmpStrobo} and \eqref{eq:DiscreteAmpSchroedi}.
        The expression we obtain for $\FDA_n$ is a sum of exponential functions in $n$, $\FDA_n = f(n)$.
        To compare it with the non-Hermitian data, we plotted $\FDP^\PSI(t) = 4 \abs{\PSI(t)}/\tau$ and $\FDP^\FDA(t) \approx \abs{\FDA_{t/\tau}}^2/\tau = \abs{f(t/\tau)}^2/\tau$ in Fig.~\ref{fig:FDP}.
        This is the interpolation that was mentioned in the caption of Fig.~\ref{fig:FDP}.

      \paragraph{Moments and total detection probability}
        The previous method was used to compute the moments for the non-Hermitian approach as well.
        The poles $\mathfrak{s}_l$, as well as $u_\PSI'(\mathfrak{s}_l)$ and $v_\PSI(\mathfrak{s}_l)$ were computed symbolically.
        From Eq.~\eqref{eq:DiscreteAmpSchroedi} we found:
        \begin{align}
          \TDP^\PSI
          = &
          \tau
          \Sum{l,l'=0}{w-1}
          \frac{v_\PSI(\mathfrak{s}_l) [v_\PSI(\mathfrak{s}_{l'})]^* }{u_\PSI'(\mathfrak{s}_l) [u_\PSI'(\mathfrak{s}_{l'})]^*}
          \frac{-1}{\mathfrak{s}_l + \mathfrak{s}_{l'}^*}
          \\ \EA{T}^\PSI
          = & 
          \frac{\tau}{\TDP^\PSI}
          \Sum{l,l'=0}{w-1}
          \frac{v_\PSI(\mathfrak{s}_l) [v_\PSI(\mathfrak{s}_{l'})]^* }{u_\PSI'(\mathfrak{s}_l) [u_\PSI'(\mathfrak{s}_{l'})]^*}
          \frac{(-1)^m m!}{(\mathfrak{s}_l + \mathfrak{s}_{l'}^*)^{m+1}}
          .
        \end{align}
        
        The stroboscopic data was obtained from the quantum renewal equation \eqref{eq:QuantumRenewal} just like for the infinite line.
        The transition amplitudes are obtained numerically from the matrix form of the Hamiltonian.
        Having numerical values for $\FDA_n$, the non-normalized moments are obtained via $\EA{T^m}^\FDA \TDP^\FDA \ApproxEq \tau^m \sSum{n=1}{N} n^m \abs{\FDA_n}^2$.
        $N$ was chosen such that the last summand is small compared to the sum.

    \subsection{Random Hamiltonian}
      A slightly different approach was chosen for the random Hamiltonian $\Ham_R$.
      Here, a complete symbolic computation of the resolvents was not possible, due to the large dimension of $\Ham_R$.
      Instead, $\Ham_R$ was  numerically diagonalized.
      Then it was renormalized via:
      \begin{equation}
        \Ham_R
        \to
        \frac{4\gamma}{E_\text{max} - E_\text{min}} \qty[ \Ham_R - \frac{E_\text{max} + E_\text{min}}{2} \Id_{32} ]
        ,
      \label{eq:}
      \end{equation}
      so that its eigenvalues lie in $[-2\gamma,2\gamma]$.

      The eigensystem contained all energy levels $E_l$ and was also used to find the overlaps $p_l$ and $q_l$.
      Having the overlaps and the energy levels, we computed the resolvents $u_\PSI(s)$($v_\PSI(s)$) semi-symbolically in Mathematica.
      This allowed us to find the poles $\mathfrak{s}_l$ and from them $\PSI(t)$ as well as all moments.

      The stroboscopic pdf, as well as the stroboscopic moments were obtained from the renewal equation.
      To achieve this, the transition and return amplitudes were computed from:
      \begin{equation}
        \mel{\PsiDet}{\TEO(n\tau)}{\PsiIn} = \Sum{l}{} p_l q_l e^{- i n\frac{\tau E_l}{\hbar}}
        ,
      \label{eq:}
      \end{equation}
      and similar for $\ev*{\TEO(n\tau)}{\PsiDet}$.
      All required quantities were given by the eigensystem of $\Ham_R$.
      The non-normalized moments were obtained as described in the last section with $N=10000$.

      However, especially for small $\tau$, convergence was an issue.
      The convergence rate of the sum $\sSum{n=1}{N} f(n) \abs{\FDA_n}^2$ is given by the largest modulus of the poles $\max{\abs{\mathfrak{z}_l}}$.
      This in turn is controlled by the magnitude of the overlaps $p_l$ and the distance between adjacent energy levels $\Delta E_l = \abs{E_{l+1} - E_l}$.
      Since we did not want to choose a much larger $N$, we picked a realization of $\Ham_R$ for Fig.~\ref{fig:RandStat} such that $N^* := \Max{\min p_l}{\min \Delta E_l^2}$ was smaller than $1000$. 
      We estimated that roughly $20\%$ of all matrices from the Gaussian unitary ensembles fall in this class.
      The results are qualitatively the same for every matrix of the ensemble.
      For matrices with a large $N^*$ (and summation with a fixed $N$) there will be a more severe dip in the stroboscopic data for small $\tau$.
      For these one would need to increase $N$ and wait much longer to obtain satisfying graphs.

\end{document}